\documentclass[10pt,conference]{IEEEtran} 
\IEEEoverridecommandlockouts
\usepackage{cite}
\usepackage{amsmath,amssymb,amsfonts}
\usepackage{algorithmic}
\usepackage{graphicx}
\usepackage{textcomp}
\usepackage{xcolor}
\usepackage{comment}

\usepackage{amssymb}
\usepackage{amsmath, amsfonts}

\usepackage{graphicx}
\usepackage{textcomp}
\usepackage{xcolor}
\usepackage{hyperref}
\usepackage{cleveref}
\usepackage{xspace}
\usepackage{paralist}
\usepackage{multirow}
\usepackage{subcaption}
\usepackage{tikz}
\usepackage{courier}
\usepackage{listings} %code highlighter
\usepackage{array}
\usepackage{graphics}
\usepackage{lipsum}
\usepackage{arydshln}
\usepackage{booktabs}
\usepackage{fancyhdr}

\usepackage{todonotes}
\setuptodonotes{inline, color=blue!30}
\definecolor{Gray}{gray}{0.3}
\tikzstyle{mybox} = [draw=black, very thick, rectangle, rounded corners, inner ysep=5pt, inner xsep=5pt, fill=gray!20]
\newcommand{\xyz}[2]{
    \smallskip
    \noindent
    \begin{tikzpicture}
        \node [mybox] (box){%
        \centering
        \begin{minipage}{.97\columnwidth}
        \fontsize{8.8}{10}\selectfont
        \textbf{RQ #1}. #2
        \end{minipage}
        };
    \end{tikzpicture}%
}

\usepackage{enumitem}
\setlist[enumerate]{leftmargin=1.2em)}
\setlist[itemize]{leftmargin=1.2em}

\def\BibTeX{{\rm B\kern-.05em{\sc i\kern-.025em b}\kern-.08em
    T\kern-.1667em\lower.7ex\hbox{E}\kern-.125emX}}
\begin{document}

%\title{Can LLMs Substitute for Human Study Subjects in Software Engineering?}%\\

\title{Can LLMs Replace Manual Annotation of\\ Software Engineering Artifacts?}

%\title{Can LLMs Replace Manual Annotation in Software Engineering Data?}

%\title{Can LLMs Replace Manual Annotation in Software Engineering Research?}

%\title{LLMs (Sometimes) are Few-Shot Data Annotators  of Software Engineering Artifacts}

%\thanks{We  acknowledge the Intelligence Advanced Research Projects Agency (IARPA) under contract
%W911NF20C0038 for partial support of this work, as well as the National
%Science Foundation, under CISE SHF MEDIUM 2107592. 
%Our conclusions do not necessarily reflect the position or the policy of
%our sponsors and no official endorsement should be inferred.}
%}

%\begin{comment}

\author{
\IEEEauthorblockN{Toufique Ahmed\IEEEauthorrefmark{1}\IEEEauthorrefmark{2}\textsuperscript{\textsection}, Premkumar Devanbu\IEEEauthorrefmark{1}, Christoph Treude\IEEEauthorrefmark{3} Michael Pradel\IEEEauthorrefmark{4}}
\IEEEauthorblockA{\IEEEauthorrefmark{1}\textit{University of California, Davis, USA}}
\IEEEauthorblockA{\IEEEauthorrefmark{2}\textit{IBM Research, Yorktown Heights, New York, USA}}
\IEEEauthorblockA{\IEEEauthorrefmark{3}\textit{Singapore Management University, Singapore}}
\IEEEauthorblockA{\IEEEauthorrefmark{4}\textit{University of Stuttgart, Germany}}
}

%\end{comment}

\maketitle

\begingroup\renewcommand\thefootnote{\textsection}
\footnotetext{Majority of the work was done when the author was a postdoctoral scholar at UC Davis.}
\endgroup

\begin{abstract}
Experimental evaluations of software engineering innovations, e.g., tools and processes,  often include \emph{human-subject studies} as a component of a multi-pronged strategy to obtain greater generalizability of the findings.
However, human-subject studies in our field are challenging, due to the cost and difficulty of finding and employing suitable subjects, ideally, professional programmers with varying degrees of experience.
Meanwhile, large language models (LLMs) have recently started to demonstrate human-level performance in several areas.
This paper explores the possibility of substituting costly human subjects with much cheaper LLM queries in evaluations of code and code-related artifacts.
We study this idea by applying six state-of-the-art LLMs to ten annotation tasks from five datasets created by prior work, such as judging the accuracy of a natural language summary of a method or deciding whether a code change fixes a static analysis warning.
Our results show that replacing some human annotation effort with LLMs can produce inter-rater agreements equal or close to human-rater agreement.
To help decide when and how to use LLMs in human-subject studies, we propose model-model agreement as a predictor of whether a given task is suitable for LLMs at all, and model confidence as a means to select specific samples where LLMs can safely replace human annotators.
Overall, our work is the first step toward mixed human-LLM evaluations in software engineering.
\end{abstract}


%\michael{Should we add something like ``in Software Engineering'' to the title? We're not the first to address the more general question.}

\section{Introduction}

Given that developer effectiveness heavily depends on good tools
and processes, researchers constantly seek more and better automation
in these areas. 
%are seeking to innovate by improving and automating parts of software engineering.
To mention just a few examples, there are now many research efforts~\cite{zhu2019automatic,zhang2022survey} aimed, \emph{e.g., } at automated code summarization, (i.e., techniques that generate a natural language summary for a given piece of code), at detecting bugs and other issues in a program~\cite{sridhara2010towards}, and at determining whether a warning produced by a static analysis tool is actually worth addressing~\cite{kang2022detecting}.

But the value of such innovations ultimately is dependent
on human judgment and practice. 
For example, a developer might read a generated English summary of a method, and then decide whether and how to use that method.
As another example, a developer might look at a warning produced by a static analysis tool, and then decide whether to fix it or not.
Because measuring human usefulness is difficult, many research efforts rely on proxy metrics.
For the example of code summarization, a summary might rate a reasonable BLEU score, indicating at least some level of similarity to the original, human-written summary.
However, even given a summary with a high BLEU score, can we be sure that it is clear, precise, relevant to the input code, and human-comprehensible? 

To  answer such questions, %determine whether the output of a novel research result satisfies such desiderata, 
researchers often use %strongly advocate for, and adopt 
human subject-based evaluations, \emph{e.g.,} when evaluating code summarization techniques~\cite{haque2022semantic,roy2021reassessing}. 
%Naturally, this evaluation approach is not just for tools that 
Besides code summarization, such evaluations are becoming increasingly common in many different settings; indeed 
in most cases where a tool is used to automate some aspect of software development,
%in most settings where some aspect of software production is automated using a supporting tool, whether software artifacts are being generated or analyzed, 
the output from the tool will be used by a human.
Since the content and form of the output may affect how well
a developer can use it, a human-subject evaluation is often vital.
%the ability of the human to properly use the output, it may be sometimes necessary to evaluate the tool using a human-subject study.

But human-subject studies in software engineering are \emph{costly}. For external validity,
such studies demand both a representative sample of developers and 
a representative sample of relevant artifact samples. Each sample may need to be evaluated by multiple humans, to get more stable results. For example, Haque et al.~\cite{haque2022semantic}, hired professional developers, at \$60 per hour, to rate a total of 420 human and machine-generated code summaries, with three ratings for
each sample to gauge inter-rater agreements.
Due to the high costs, such evaluations
are sometimes done using free or low-cost participants, such as students, which carries risks %obviously carries some threats
of the results not generalizing to professional developers.
Yet, even performing a human-subject study with students can be quite time-consuming.
%, and also costly, if the students are paid
% (in many settings, at least the legal minimum wage must be paid). Ideally, one should prefer
% to do such studies using professional programmers. Given
% \prem{fix the numbers}
% the per-sample time of $xx$ minutes reported in ~\cite{reviewer2}, and a reasonable programmer wage as reported in the website~\cite{reviewer2}, we end up
% a total cost of around \$ $yy$,000 for the study. Certainly the high cost of programming labour amplifies the cost of such experiments in software engineering.
Thus, the main motivation for researching new tools and techniques in software engineering, \emph{viz.} the high cost of developers, is also a hurdle to  the proper evaluation of such tools and techniques. 

Given the ability of advanced large language models (LLMs), 
%and their ability 
to rival human performance in a range of tasks~\cite{achiam2023gpt,reid2024gemini, claude2024}, the question naturally arises:
\emph{Can we use LLMs to reduce the cost of human-subject studies in software engineering?}
If LLMs can help, even if only  partially, 
%for a subset of the effort involved in such studies, 
this could impact the practice of evaluation studies in software engineering.
Thus, we, ask: % seek to understand 
%Moreover, improving our understanding of 
When, and how, can  human subject responses be safely replaced by LLMs, in a mixed human-LLM evaluation scenario?  
%is an important first step toward such mixed human-LLM evaluations.

This question has arisen in other disciplines, \emph{e.g.,} psychology, linguistics, and medicine~\cite{aher2023using,futrell2019neural,goel2023llms,bavaresco2024llms}.
There are several reasons why the software engineering domain is particularly interesting for this exploration: First, relative cost: LLMs can run hundreds of queries for the cost of using a single human subject for an hour.
Second, LLMs now perform well on a wide range of software engineering tasks, 
%are well-trained on software
%artifacts, both code and code-related natural language, 
%have proven value for  a wide range of software engineering tasks, 
and are routinely used in % the software 
industry~\cite{dunay2024multi,peng2023impact,achiam2023gpt,team2023gemini}.
Third, %advances in prompt engineering, such as 
the impressive capabilities of LLMs at  
in-context learning~\cite{ahmed2022few,brown2020language}, suggest that with a few illustrative examples, 
LLMs can execute fairly complex software engineering tasks. Finally,  software artifacts
are arguably more complex than natural language artifacts: they can require deep knowledge of
both application domain and programming; also, they comprise multiple formal \& informal elements (\emph{e.g,} code, summary, warnings).

This paper studies the possibility of substituting human subjects with LLMs when annotating  software engineering artifacts.   
We study ten tasks from five datasets created by prior work, covering research ranging from requirements engineering %over code summarization 
to reasoning about static analysis warnings.
We apply six state-of-the-art LLMs, including both open- and closed-source models, to these tasks, and then compare the LLMs' responses to those from human subjects.

We focus on tasks whose subjective, 
nuanced and context-dependent nature
have traditionally required human judgment. 
%subjective judgment is essential because these tasks traditionally rely on human evaluators due to their nuanced and context-dependent nature.
%for a specific reason! Simpler tasks (like bug report classification) can already be done using older generations of ML systems. 
This subjectivity is evident in the fact prior studies 
reveal considerable inter-rater \emph{dis}agreement. % is not rare. 
%We have observed that models like GPT-4 perform well on the samples where all the reviewers provided the same annotations. 
Figure~\ref{fig:example} shows two examples from code summarization, where human subjects are asked to rate the accuracy of generated summaries.
In Figure~\ref{fig:sub1}, all the three raters and the GPT-4 model rated ``Strongly agree''.
However, other examples are inherently hard for humans, such as Figure~\ref{fig:sub2}, where the three raters give different ratings: ``Agree'', ``Strongly Disagree'', and ``Disagree'', and GPT-4 agrees with the first rater.
Our study asks: \emph{Can powerful LLMs (partially) substitute for humans on inherently subjective annotation tasks?}. Our findings are: 

\begin{figure}[t]
    \centering
    % First subfigure on top
    \begin{subfigure}{\columnwidth}
        \centering
        \includegraphics[width=0.95\columnwidth]{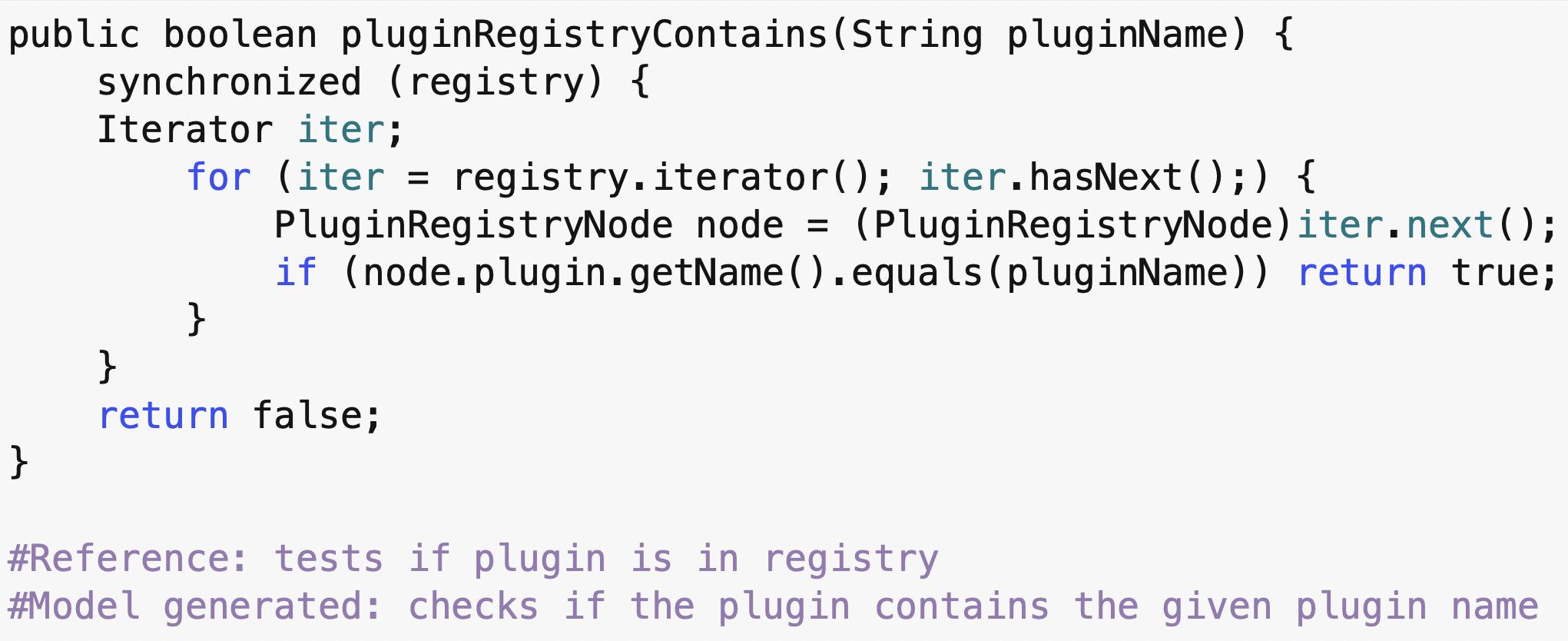}
        \caption{Sample where human raters agreed}
        \label{fig:sub1}
    \end{subfigure}

    \vspace{.3cm}

    % Second subfigure below
    \begin{subfigure}{\columnwidth}
        \centering
        \includegraphics[width=0.95\columnwidth]{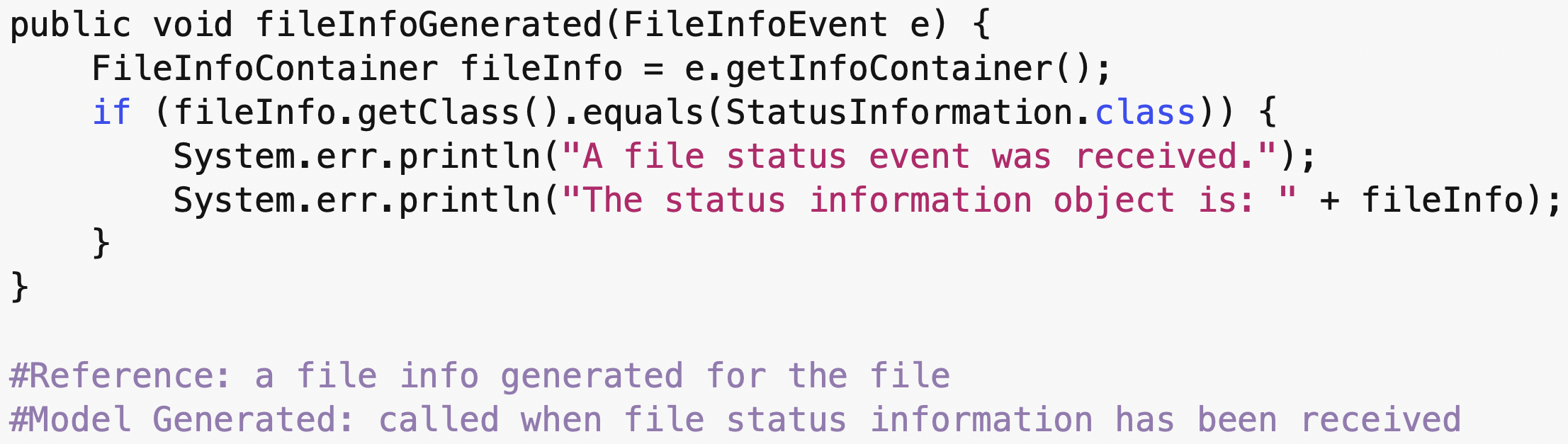}
        \caption{A sample where humans disagreed}
        \label{fig:sub2}
    \end{subfigure}
 \vspace{.1cm}
    \caption{Annotation tasks with different difficulty level.}
    \label{fig:example}
\end{figure}

\begin{itemize}
  \item On some tasks, LLMs agree with human subjects about as much as human subjects agree with each other.
  For example, when judging whether an identifier name and the value it refers to are consistent~\cite{patra2022nalin}, we observe a mean human-human inter-rater agreement of 0.52, and a mean human-model agreement of 0.49.
  This results suggests that LLMs can sometimes be a meaningful replacement for humans on \emph{some} annotation tasks. But when, exactly? 
  \item To help researchers decide whether an LLM may be suitable for a specific annotation task, we investigate the correlation between model-model agreement, which can be measured
  programmatically, and the ability of LLMs to meaningfully replace humans.
  We find model-model agreement strongly correlates
  with human-model agreement, 
  %(Spearman correlation of 0.65), 
  suggesting that model-model agreement can  be used
%  serve as a relatively cheap-to-obtain proxy 
to decide whether to involve LLMs in an evaluation.
  \item Some examples are easier to annotate than others. For a given example, can an LLM substitute for a human?
  We find that the LLM's confidence (its output probability) helps select examples that do not necessarily require human annotators.
  For example, for the task of judging whether a code snippet and its summary are similar, delegating 50\% of the effort performed by one annotator to an LLM (selected based on LLM confidence), does not lead to a statistically significant change of the overall inter-rater agreement.
\end{itemize}

\medskip \noindent
In summary, we make the following contributions. 
\begin{itemize}
\item We offer a first study as to whether LLMs can replace human annotation on software engineering artifacts.
\item We present a methodology designed to answer this question and empirical results from applying six LLMs to ten tasks addressed by humans in prior work.
\item We propose a method for selecting which \emph{tasks} are suitable for an LLM-augmented evaluation and which \emph{samples} can be safely delegated to an LLM.
\end{itemize}

%, and also automatically choosing/ranking static analysis warnings for  possible repair. Many of these efforts use machine-learned models, some not. 

%test citation~\cite{haque2022semantic}
\label{intro}
\section{Background}

In software engineering research,  
%evaluations of tools and techniques often rely on qualitative annotations of software artifacts (\emph{e.g.,} code, documentation, bug reports, or newly generated research artifacts~\cite{easterbrook2008selecting}).
human-subject annotations provide insight into the 
potential impact of innovations on %and their potential impact on 
productivity or quality~\cite{easterbrook2008selecting}; but 
the nuanced, subjective nature of these evaluations  often
%Traditionally, many such evaluations involve human subjects, which 
requires multiple human raters to ensure consistency and reliability in the data~\cite{morse2002verification}. 

Inter-rater reliability  provides a vital gauge for the objectivity of the annotations~\cite{lombard2002content}. It is commonly measured using statistical metrics such as Cohen's $\kappa$~\cite{cohen1960coefficient} or Krippendorff's $\alpha$~\cite{krippendorff2018content}, which measure agreement among annotators by considering the possibility of pure-chance. %agreement occurring by chance.
Higher values suggest a strong, reliable, and robust agreement among annotators.
%suggesting that the annotations are reliable and that the findings are robust.

Annotations based on a single annotator carry a risk  of
%there is a significant risk that the findings may reflect 
personal biases and subjective interpretations,  and lead
to inaccurate, non-generalizable
%which could lead to inaccuracies and limit the generalizability of 
results~\cite{finlay2002outing}. In contrast, multiple annotators working independently can reveal an inherently shared understanding, which indicates that the annotations accurately capture the underlying phenomena being studied. However, achieving high inter-rater reliability is effortful, time-consuming, and challenging; 
human developer time is costly, and they often disagree!
%, often involving substantial manual work. 
%This challenge is compounded by the variability in human interpretation, which can lead to inconsistencies in the data.

LLMs, since they are trained on very large corpora,  
could help these challenges by potentially providing more consistent annotations which normatively reflect ``most" developers. 
They %Properly trained LLMs 
could reduce the biases introduced by human annotators,
%and help afford the scaling of qualitative analysis to much larger datasets. 
LLM-based automation of annotations could 
%open up new research opportunities by 
make it feasible to analyze datasets that were previously too large or complex to annotate manually. In addition, software engineering data is dynamic and constantly evolving, requiring tools that can quickly adapt to new information. LLMs, with their ability to learn from and adapt to new data, could provide up-to-date annotations, complementing the work of human annotators.

To effectively incorporate LLMs into qualitative annotation, it is essential to establish a structured process. This includes deciding when and how to use LLMs, integrating them into existing workflows, distributing tasks between humans and LLMs, and determining the degree of manual work required. This paper addresses these considerations through four research questions (See \autoref{sec:RQs}), focusing on annotations with pre-existing categories derived from related work or developed in an initial research phase. Future work will explore the potential of LLMs for annotating qualitative data without pre-existing categories (i.e., open coding) within a software engineering context.
\section{Methodology}

%\michael{Move RQs into this section?}

We now describe the tasks, datasets, the models under consideration, our research questions, and the methodology to answer these questions.  
%We tried to use the exact same instructions for the models that were used in the original study.
% In each case, we describe 
% first the original taks (with human subjects) and our approach to challenging an LLM
% with the same task, and measuring the results. 

\subsection{Tasks \& Datasets}

We select five datasets from previous work, which together include 10 human-annotation tasks.
For some datasets, multiple tasks were assigned to the raters. For example, in the code summarization dataset, raters were given four tasks: rating accuracy, adequacy, similarity, and conciseness. Similarly, for the semantic similarity dataset, raters were assigned three different rating tasks. This is why we have five datasets or major tasks but a total of 10 human-annotation tasks. 
In general, we ask LLMs to perform
annotations previously done by humans; we give the
LLMs the same instructions as given to humans.
The datasets and tasks are selected to represent a diverse range of software engineering research.

\noindent{\underline{\em Automatic Code Summarization}}
This is an active area, aiming to generate helpful summaries of code~\cite{sridhara2010towards}.
One survey-based study reports that 80\% of respondents found code comment generation tools useful~\cite{hu2022practitioners}. 78\% agree that these tools help them understand the source code, especially helping code readability in  projects with few comments. %and thus help improve code readability.
Several metrics are used to evaluate code summary quality. All metrics aim to indirectly measure \emph{human perception of quality}, but have limitations. % they all aim to correlate with human perceptions. 
But what criteria are important to evaluate human perception? 
Haque et al.~\cite{haque2022semantic} considered four criteria: accuracy, adequacy, conciseness, and similarity—that should be considered when evaluating code summary quality.

Haque et al.~\cite{haque2022semantic} recruited experienced programmers via Upwork, paying them USD60/hr (the applicable market rate). The programmers were presented with 210 functions (from LeClair \emph{et al}~\cite{leclair2019neural}) along with associated human-generated, 
and model-generated summaries (420 in total).
Subjects responded to four questions, related to each of the criteria above. Each participant rated a subset of the 420 samples\footnote{https://github.com/similarityMetrics/similarityMetrics}, while ensuring that each sample was rated by at least three people.

To elicit  LLM responses, we prompt the model with the same guidelines and questions as in the original study.
That is, the LLM receives the functions and associated comments, and is tasked to rate its agreement with four statements:
\begin{itemize}
    \item Independent of other factors, I feel that the summary is accurate.
    \item The summary contains all the important information, and that can help the understanding of the method.
    %\underline{\bf FIX THIS and NEXT TO BE ACCURATE!!}
    \item  The summary contains only necessary information.\footnote{As noted in Truong \emph{et al}~\cite{truong2023language}, LLMs do not handle negation accurately; so this and the above instruction were slightly modified to remove negation, without changing meaning. \label{ref:neg}}
    \item These two comments are similar.
\end{itemize}
There are four options: `Strongly agree', `Agree', `Disagree', and `Strongly disagree'. To enhance model performance and maintain a specific output format, we use few-shot learning~\cite{brown2020language,ahmed2022few}:
%instead of zero-shot learning. 
we prompt the model with a few illustrative query-response pairs and expect the model to respond to the ``test'' query of our interest. 
Prompt-size constraints limited us to three shots for most of the experiments.
%, constrained by the prompt token limitations of multiple models.

%We found that the models 
%erred consistently in one direction, specifically for
%with adequacy and conciseness: because of the way the questions were asked. Just for these two questions, a lower score (Disagree/Strongly disagree) reflects
%higher rating of comments. We simply eliminated the negative sense in both the response and the answer so that adequate and concise comments reflect higher scores. 
%\prem{Put the actual wording we used (already double-negated) in the text above, and explanation in footnote} In this paper, for these two questions, we will report the results with the double negation.

\smallskip
\noindent{\underline{\em Name-Value Inconsistencies }}
Variable names are crucial for code understanding~\cite{patra2022nalin}. If a variable name does not match the value stored in it, we have an undesirable \emph{``name-value inconsistency"}.  
Patra and Pradel~\cite{patra2022nalin} user study had 11 participants, and created a dataset  evaluating name-value consistency\footnote{https://github.com/sola-st/Nalin}. This dataset includes 40 samples rated by all 11 raters. The raters assigned a Likert score from 1 to 5, where 1 indicates difficulty in understanding (or the presence of name-value inconsistency), and 5 indicates ease of understanding.
Suppose there is a variable named {\small\tt name} to which a float, 2.5, is assigned. This is confusing for developers, since the expected value is a string. All  annotators rated it as difficult to understand. On the contrary, another variable {\small\tt done}, assigned the boolean value False, is easy to understand and was rated as easy to understand by most annotators.
We challenge the LLM to perform the same tasks, again with a few-shot  prompt. %\prem{instructions? prompting?}

\smallskip
\noindent{\underline{\em Causality}}
Causal knowledge supports reasoning about requirements dependencies,
particularly for tasks  such as test-case construction. Causal relations (e.g., `If event 1, then event 2') are common in system behavior. However, extracting causality from natural language documents is difficult. Fischbach et al.~\cite{fischbach2021automatic} introduced a dataset, built using 6 raters to develop and evaluate a tool for extracting causal dependencies.\footnote{https://github.com/fischJan/CiRA} The dataset is quite large, with over 10,000 samples. We randomly select 1,000 samples where at least two raters assessed each sentence. The ratings are binary: 1 indicates the presence of a causal relation, and 0 indicates its absence.
We elicit such annotations from  LLM 
%o perform the same annotation task as the original participants 
using a few-shot prompt.
%instead of zero-shot learning.
%\prem{instructions? prompting?}

\smallskip
\noindent{\underline{\em Semantic Similarity}}
Functions are often re-implemented several times, with similar functionality, 
in software projects.
However, semantically-similar functions may be implemented in different ways. Finding and merging such functions could reduce maintenance costs.
%by allowing the reuse of test cases and assisting in fixing various bugs. 
Kamp et al.~\cite{kamp2019sesame} applied text similarity measures to JavaDoc comments mined from 11 open-source repositories and then manually classified a selection of 857 pairs to create a realistic dataset.\footnote{https://github.com/FAU-Inf2/sesame}
There are three annotation tasks related to functional similarity: goals, operations, and effects. Like for code summarization, each sample was rated by three different people from a pool of eight raters.
The original dataset includes some metadata and URLs to the functions but does not contain the function bodies. We retrieve the function bodies and found both bodies for 786 pairs, so we conduct our evaluation on these samples. 
For each of these pairs, we query LLMs to perform the same three annotation tasks as in the original study.

\smallskip
\noindent{\underline{\em Static Analysis Warnings}}
Automatic static analysis tools, (\emph{e.g.} FindBugs), can find coding errors, but can raise false alarms. Pruning false alarms and presenting only (or mostly) actionable warnings to developers can be beneficial.
A dataset by Kang \emph{et al}.~\cite{kang2022detecting} provides human annotations on the task of determining whether a particular code change effectively addresses a static analysis warning.
The dataset has 1,306 samples that cover different warning categories.\footnote{https://github.com/soarsmu/SA\_retrospective/tree/main}
Two annotators assigned each sample one of three labels: open, closed, or unknown. A static analysis warning is considered ``closed'' if it was reported in a previous software revision but not in a subsequent reference revision, indicating that the problematic code was altered or removed. Conversely, a warning is labeled as ``open'' if it appears in both the current testing revision and a later reference revision, suggesting that the warning was not actionable or it was ignored by the developers. Lastly, a warning is marked as ``unknown'' if the file containing the warning was deleted or modified in a (possibly unrelated) way in the reference revision, making it difficult to confirm whether the warning was actionable. %, as the removal or modification of the file could have been unrelated to the warning itself.
We use diff to present the changes made to the repository.
Given the extensive metadata, the prompts for this problem are the longest. Due to rate limitations imposed by some models (e.g., Claude and Gemini) and the need to manage costs, we use 200 samples, chosen uniformly at random, for our experiments.
For each of these samples, we ask LLMs to perform the annotation task.

\bigskip
\noindent \emph{Detailed prompts (including guidelines and few-shot samples) for each dataset are provided in the supplementary material}.

\subsection{Models under Consideration}
We use both closed and open models. Among closed models, we chose GPT-4, Claude-3.5-Sonnet, and Gemini-1.5-Pro, which are all the best models from their respective families or organizations at the time of the work. We also used the GPT-3.5 model. From the open models, we chose Llama3 (70B) and Mixtral (8x22B), which are also the latest models from the open-model family. 
%All the models are highly capable

\subsection{Research Questions}
\label{sec:RQs}
%In this paper, the primary 
We seek to study whether LLM prompting can substitute for human annotation labour, for artifact annotation tasks in SE. 
Specifically, we study how  annotator (inter-rater) agreement  changes when replacing human annotations with automated LLM annotations. In the first research question, we examine all three categories of inter-rater agreement: human-human, human-model, and model-model. Specifically, if \emph{human-model} inter-rater agreement is similar to that of \emph{human-human} agreement, it may be possible to substitute (some) human ratings with  model ratings. We will also examine the inter-model agreement and how it changes across datasets.

\smallskip
\xyz{1}{When humans are replaced by LLMs to provide answers for ``human-rater'' questions in software engineering research, what level of agreement is observed between humans and models?}
% should we do eachhuman-LLM and LLM-allhumans agreement. 

%It appears to be useful in some tasks, but not in others; so we need a process for deciding if we can susbtitute, and when we should do that (inter-model agreement, and some human samples to
%justify polarity etc) 

%Right now, it appears that inter-model agreement is a good indicator of whether model agrees with human. SO 1) find inter-model agreement; 2) if good, pick the best model you can afford. 

%Now, is this above rule good for {\bf all} models? we only did 4.
%should do more (k $>t$ 4) and see if this rule holds for all $^kC_4$ %choices? 

In Section~\ref{res_rq1} we discuss the findings of human-human, human-model and model-model inter-rater agreement. Depending on the dataset and annotation task, we find varying levels of agreement. This raises the question of how one might decide whether a particular task is amenable to replacing human effort with an LLM.  
%For some datasets we observed high human-model agreement, where we can easily replace a human with a model and for others the agreement is low and we can not entirely replace a human with the model. In this research questions we will try propose a technique that will enable us to to find out whether a model is good at certain task. If not how can we utilize the model at those task. %

\xyz{2}{How can we determine if LLMs are NOT usable for a specific task?}
%Even it is useful, it may have a modest success rate. In this
%case, can we predict on a per-sample basis, whether the model's
%answer is likely to agree with the human answer?

In contrast to RQ2, where we consider basic feasibility of replacing  human effort with a model, the next question focuses on identifying \emph{specific samples} where we can replace one human rating with a model rating. %, or completely replace all human ratings with  model ratings. If the model's rating aligns with the majority of the ratings, we consider one human is replaceable. Note that for some tasks, like datasets with only two raters per sample, we consider ``majority" as equivalent to agreeing with all raters. In datasets with three raters per sample, we consider samples where model agrees with two or more raters, to be one where the model rating could substitute  for a human rating. 
Although a model may not perform well for all samples, selecting samples where models are successful can still significantly reduce human effort.

\smallskip
\xyz{3}{How can we determine if an answer from an LLM
for a specific sample is likely to be in agreement with a human answer?}
%\xyz{4}{How should one deal with the polarity in the prompt}

In case we have determined if and when to delegate parts of a human-subject study to LLMs, the final research question is about the cost benefits of doing so.

\smallskip
\xyz{4}{How much human annotation effort could be saved without sacrificing inter-rater agreement?}

\subsection{Evaluation Methodology}

\noindent{\underline{\em For RQ1}}: 
With each task, we record a) number of annotators and b) samples rated  by each annotator. Note that an annotator may not have annotated all the samples in the dataset but rather a subset.\footnote{For example, Haque \emph{et al} used six annotators, but each sample only received three annotations.} 
After identifying the samples annotated by each human, we compute the inter-rater agreement metric Krippendorff's $\alpha$ for each human-human, human-model, and model-model pair. To clarify, if a human (P1) has rated 5 samples \{s1, s2, s3, s4, and s5\}, and another human (P2) has rated 5 samples \{s3, s4, s5, s6, s7\}, we consider only \{s3, s4, s5\} for computing Krippendorff's $\alpha$ for P1 and P2. 
Other inter-rater agreement metrics exist; \emph{but  Krippendorff's $\alpha$ is applicable to any number of annotators}, each assigning one value to one unit of analysis, to incomplete (missing) data, to any number of values available for annotating a variable, and to different kinds of annotations (binary, nominal, ordinal, interval, ratio, etc.).
Given the diversity of the datasets we study, this metric allows us to uniformly apply inter-rater agreement to all datasets.

\smallskip
\noindent{\underline{\em For RQ2}}: As mentioned earlier, we compute human-human, human-model, and model-model agreement
$\alpha$. Note that model-model agreement is cheap and automated to determine. To answer this question, we examine the model-model agreements and human-model agreement. If these two are correlated, then model-model agreement can help judge the possibility of having high human-model agreement.%, thus increasing the chances of replacing humans.

%Our findings  model-model inter-rater agreements are moderately correlated with human-model agreements, which we aim to maximize. 
%To answer this question, we %followed two steps: we 
%examined the model-model agreements 

\smallskip
\noindent{\underline{\em For  RQ3}}:
To determine whether asking an LLM for a specific sample is a good idea, we use the output probability of a model's answer as a proxy for the model's confidence.
We then study the impact of replacing those human answers where an LLM gives the highest-confidence responses with an LLM answer.
To assess the impact, we investigate how the inter-rater agreement changes if we replace \emph{one} human annotator with an LLM for a specific fraction of all samples.

% To see whether an LLM's answer for a specific
% sample is likely to be in agreement with a human answer we repeatedly choose one rating randomly from a sample to replace it with an LLM output, then calculated the inter-rater agreement.
% In several cases, we observed that replacing one rating with a model does not change the inter-rater agreement at all. However, in some cases, model annotation is too different
% from human annotation, which is thus not replaceable. %\prem{following sentence not clear. what does "usually higher" mean} 

\smallskip
\noindent{\underline{\em For  RQ4}}:
We measure the human effort one could save without reducing inter-rater agreement in a statistically significant way.
Our unit of \emph{effort} here is the work required, on average, to label one sample.
Saving 100\% of the effort for one rating corresponds to getting one annotation on each sample in a dataset from an LLM.
Note that this saving typically saves at least one, and sometimes multiple human participants.
For example, if a problem requires a total of 300 annotations for 100 samples, then saving 100\% rating effort means to replace 100 annotations (which otherwise would be annotated by one or more humans).

%We rely primarily on the output probability associated with the GPT-4 response. First, we assign a label on each sample based on the output from GPT-4: if GPT-4 output agrees with the majority of the human raters, we consider that specific sample amenable to LLM annotation, and assign class ‘1’; otherwise, we assign class ‘0’. We order the samples by decreasing probability and plot the fraction of solutions of class '1'. We also calculate the ROC-AUC for all models that provide output probabilities to judge the predictive value of model probabilities. Note that for multi-subtoken outputs like 'Strong agreement,' we calculate the geometric mean of all subtoken probabilities. Unfortunately, some highly-performant models like Claude and Gemini do not provide output probabilities. 
%Also, the open-source models we tried  under-perform  in this
%setting. Thus, we mostly stick to GPT-4 for this research question. 
%As above, for some samples, we risk deviating significantly from the human answers, 
%if we were to substitute model answers for human answers. 
%it is difficult to replace all the human input for a sample, and if we do so, we must consider a certain amount of risk of arriving at misleading conclusions. 
%This is also consistent with other studies conducted in the NLP field~\cite{}.
%We discuss the detailed results in Section X.
\section{Results}
\label{sed:result}
\label{sec:result}

\subsection{RQ1: Level of Agreement between Humans and Models}
\label{res_rq1}
We now present the observed level of agreement between humans and models across different datasets.

\begin{table}[h]
%\caption{Effectiveness of multi-lingual fine-tuning for code summarization task}

\centering

\resizebox{\columnwidth}{!}{%
\setlength{\tabcolsep}{1pt}
\begin{tabular}{@{}lcccccc@{}}
\hline
\multicolumn{1}{c}{\multirow{2}{*}{Datasets}} & \multicolumn{2}{c}{\begin{tabular}[c]{@{}c@{}}Human to Human \\ Inter-rater agreement\end{tabular}} & \multicolumn{2}{c}{\begin{tabular}[c]{@{}c@{}}Human to Model \\ Inter-rater agreement\end{tabular}} & \multicolumn{2}{c}{\begin{tabular}[c]{@{}c@{}}Model to Model \\ Inter-rater agreement\end{tabular}} \\ \cline{2-7}
\multicolumn{1}{c}{}                          & Mean                                            & Median                                            & Mean                                            & Median                                            & Mean                                            & Median                                            \\ \hline
Accuracy (Code Summarization)                 & 0.38                                            & 0.44                                              & 0.48                                            & 0.48                                              & 0.76                                            & 0.78                                              \\
Adequacy (Code Summarization)                 & 0.40                                            & 0.39                                              & 0.41                                            & 0.38                                              & 0.74                                            & 0.73                                              \\
Conciseness (Code Summarization)              & 0.24                                            & 0.26                                              & 0.21                                            & 0.26                                              & 0.74                                            & 0.75                                              \\
Similarity (Code Summarization)               & 0.64                                            & 0.66                                              & 0.44                                            & 0.43                                              & 0.68                                            & 0.68                                              \\
Name-Value Inconsistencies                    & 0.52                                            & 0.52                                              & 0.49                                            & 0.48                                              & 0.66                                            & 0.67                                              \\
Causality                                     & 0.44                                            & 0.49                                              & 0.22                                            & 0.28                                              & 0.39                                            & 0.25                                              \\
Goals (Semantic Similarity)                   & 0.83                                            & 0.83                                              & 0.77                                            & 0.78                                              & 0.82                                            & 0.81                                              \\
Operations (Semantic Similarity)              & 0.74                                            & 0.77                                              & 0.67                                            & 0.68                                              & 0.77                                            & 0.79                                              \\
Effects (Semantic Similarity)                 & 0.71                                            & 0.73                                              & 0.64                                            & 0.65                                              & 0.69                                            & 0.64                                              \\
Static Analysis Warning                       & 0.80                                            & 0.80                                              & 0.15                                            & 0.23                                              & 0.12                                            & 0.09   \\ \hline                                          
\end{tabular}

}
\caption{Summary of inter-rater agreements. The three best performing models (Claude, Gemini, and GPT-4) have been used to report the agreement in this table.}
\label{tbl:sira-all}
\end{table}

\begin{figure*}[ht]  % The 'figure*' environment ensures it spans across two columns
    \centering
    \begin{subfigure}[b]{0.48\textwidth}  % Adjusted each subfigure to take up about half of the text width
        \centering
        \includegraphics[width=.85\linewidth]{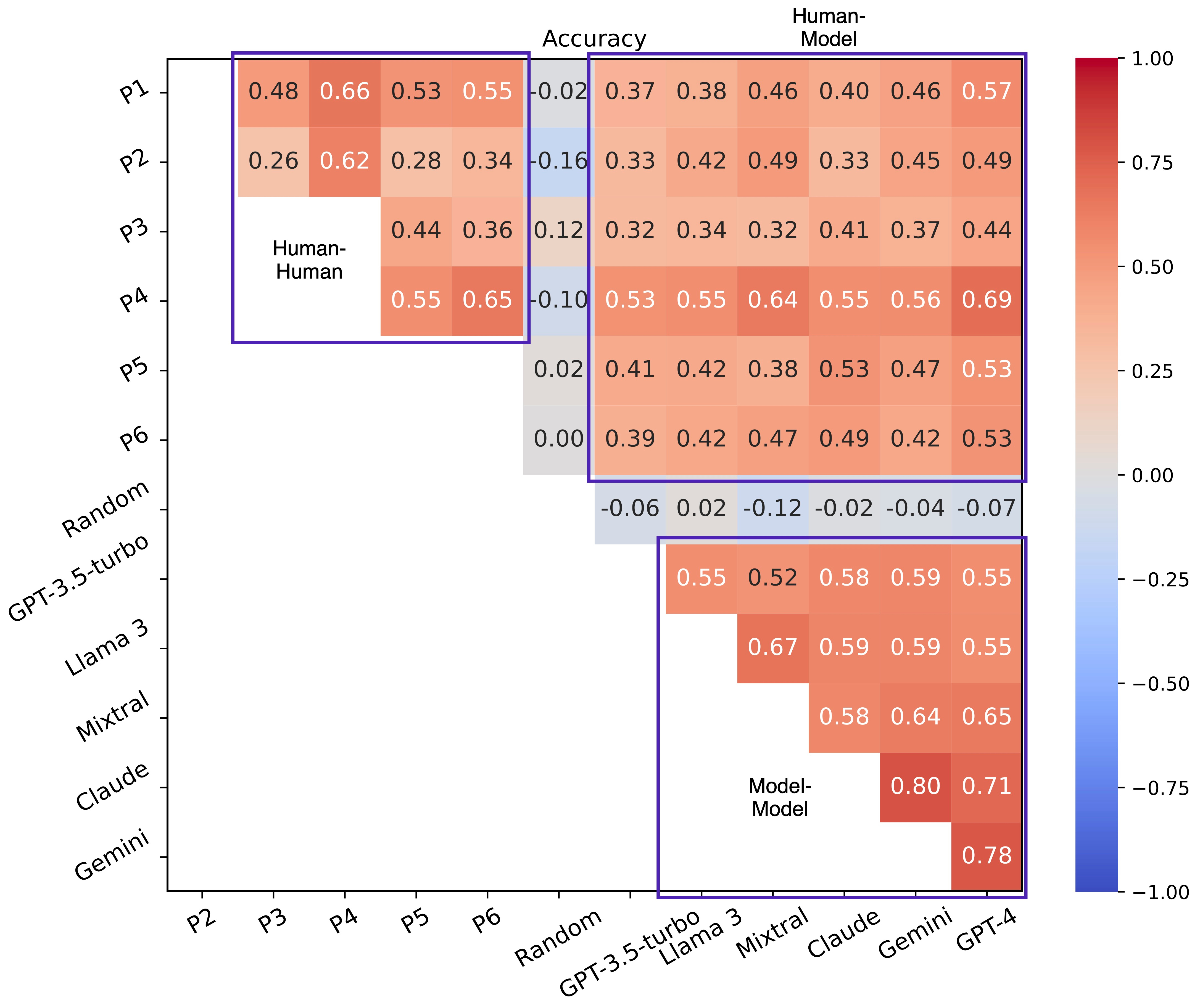}  % Scale the image to fit the subfigure width
        \caption{Accuracy}
    \end{subfigure}%
    \hfill  % Adds space between the subfigures
    \begin{subfigure}[b]{0.48\textwidth}
        \centering
        \includegraphics[width=.85\linewidth]{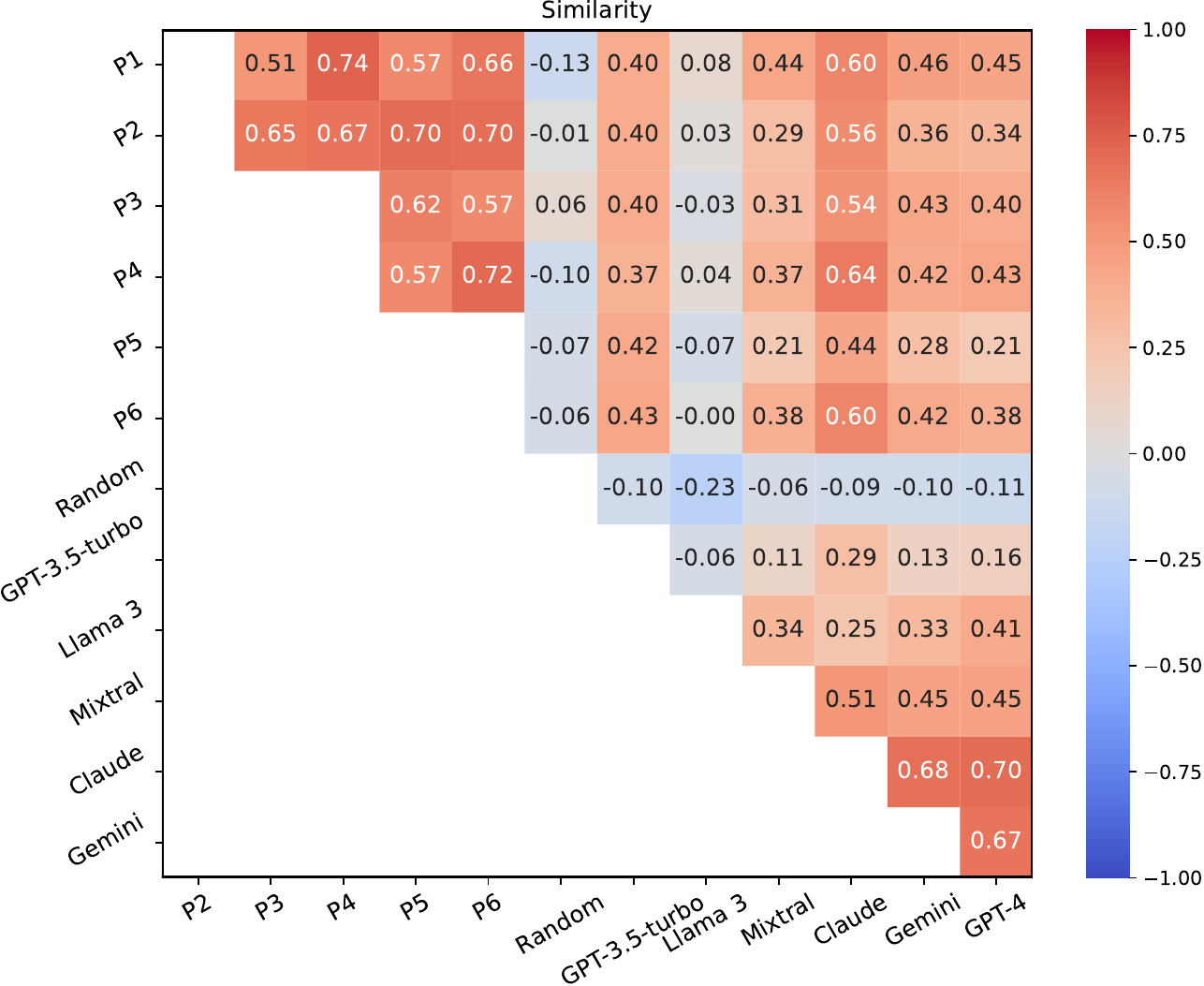}
        \caption{Similarity}
    \end{subfigure}%
    
    \caption{Inter-rater agreement (Krippendorff's $\alpha$) for code summarization accuracy and similarity. Results for adequacy and conciseness are similar (omitted due to space).}
    \label{cs-heat}
\end{figure*}

\smallskip
\noindent{\underline{\em Code Summarization}}
Figure~\ref{cs-heat} shows heatmaps of inter-rater agreement over individual ratings on experimental samples, for human-human, human-model, and model-model pairs for code summarization criteria: \emph{Accuracy, Adequacy, Conciseness}, and \emph{Similarity}. The original study used 
%employed 
six human raters. We add ratings from six different models, along with (for comparison)
an automated rater which randomly assigned labels to the samples. Figure~\ref{cs-heat}-(a) shows three zones representing three categories of inter-rater agreement, based on available data.\footnote{Some human raters did not rate
the same samples.} 

Model-model agreement is high, for all criteria, especially for the three large models (GPT-4, Gemini, and Claude). Table~\ref{tbl:sira-all} indicates that the mean Krippendorff’s $\alpha$ is 0.68-0.76. Second, we see that
human-model and human-human agreements are in similar ranges, 0.24-0.40 and 0.21-0.48 for the first three categories. Because of the way models are trained,
we can expect powerful models to tend to reflect the ``majority
opinion" they learn from the training corpus. 

For \emph{similarity}, human-human agreement is higher than human-model agreement. Although smaller models are also useful, they show some bias towards specific answers. For example, Llama-3 consistently chose between ``Strongly agree'' and ``Strongly disagree'', ignoring the options ``Agree'' and ``Disagree''. Similarly, human rater ``P2'' was not particularly satisfied with the \emph{conciseness} of the program, resulting in lower inter-rater agreement with all humans and models.

To summarize, for all criteria, we observed generally similar human-human and human-model inter-rater agreements, with a few exceptions.

\begin{figure}[!t]
    \centering
    \includegraphics[width=0.85\columnwidth]{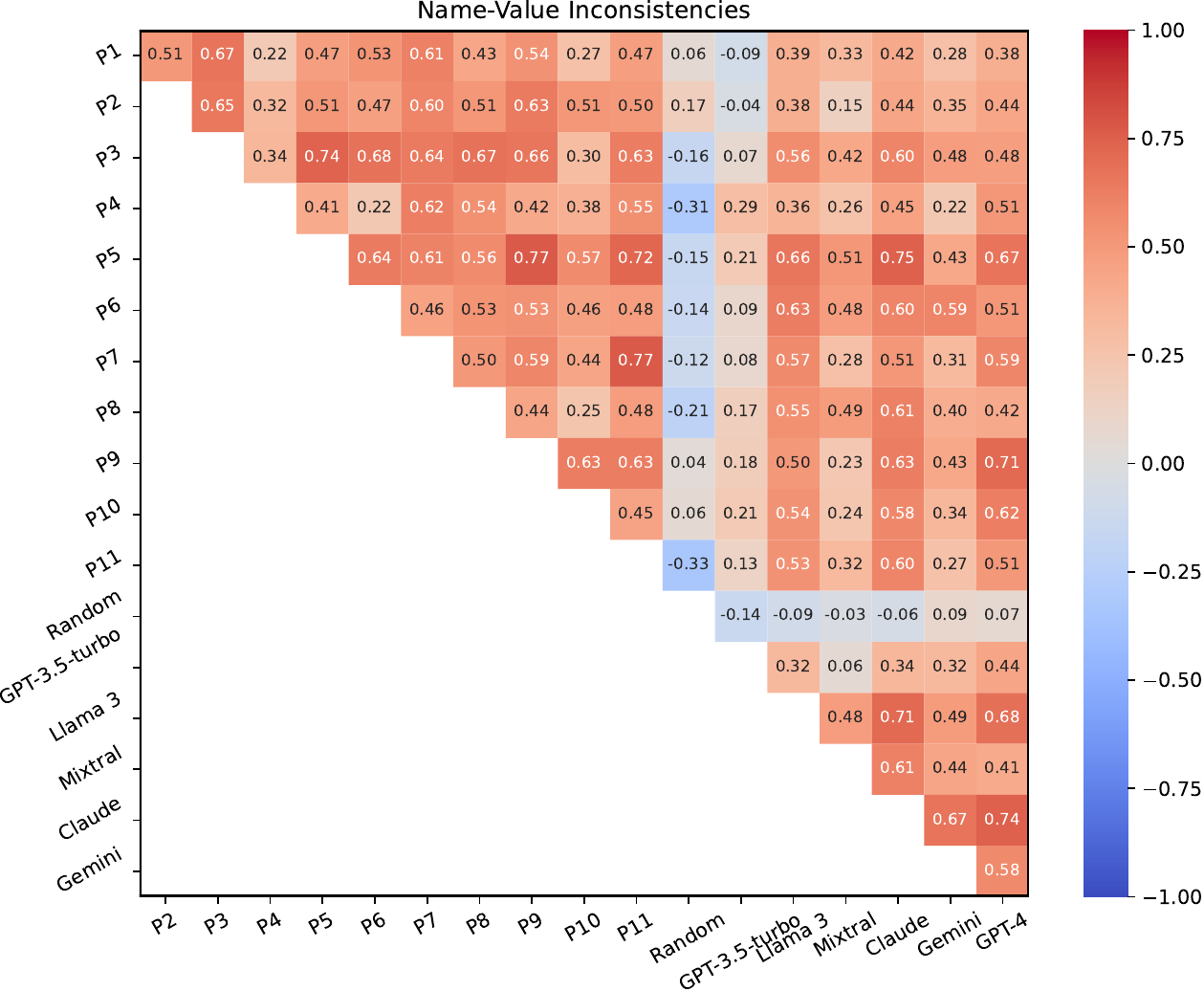}
    \caption{Inter-rater agreement for name-value inconsistencies.}
   \label{fig:nalin}
\end{figure}

\smallskip
\noindent{\underline{\em Name-Value Inconsistencies }}
We find that human-model and human-human agreement 
are quite similar for this task (mean Krippendorff’s $\alpha$ 0.49 vs. 0.52). Figure~\ref{fig:nalin} shows a heat-map for all 11 humans, 6 models, and a random rater. We have very high values in all three zones: human-human, human-model, and model-model. Here, too, we find  higher agreement between the top 3 models (mean Krippendorff’s $\alpha$ 0.66). Note that we will consider only the top three models for reporting Krippendorff’s $\alpha$ from here on because we found that only these three models do not show bias or preference towards any specific answer.

\begin{figure}[!t]
    \centering
    \includegraphics[width=0.85\columnwidth]{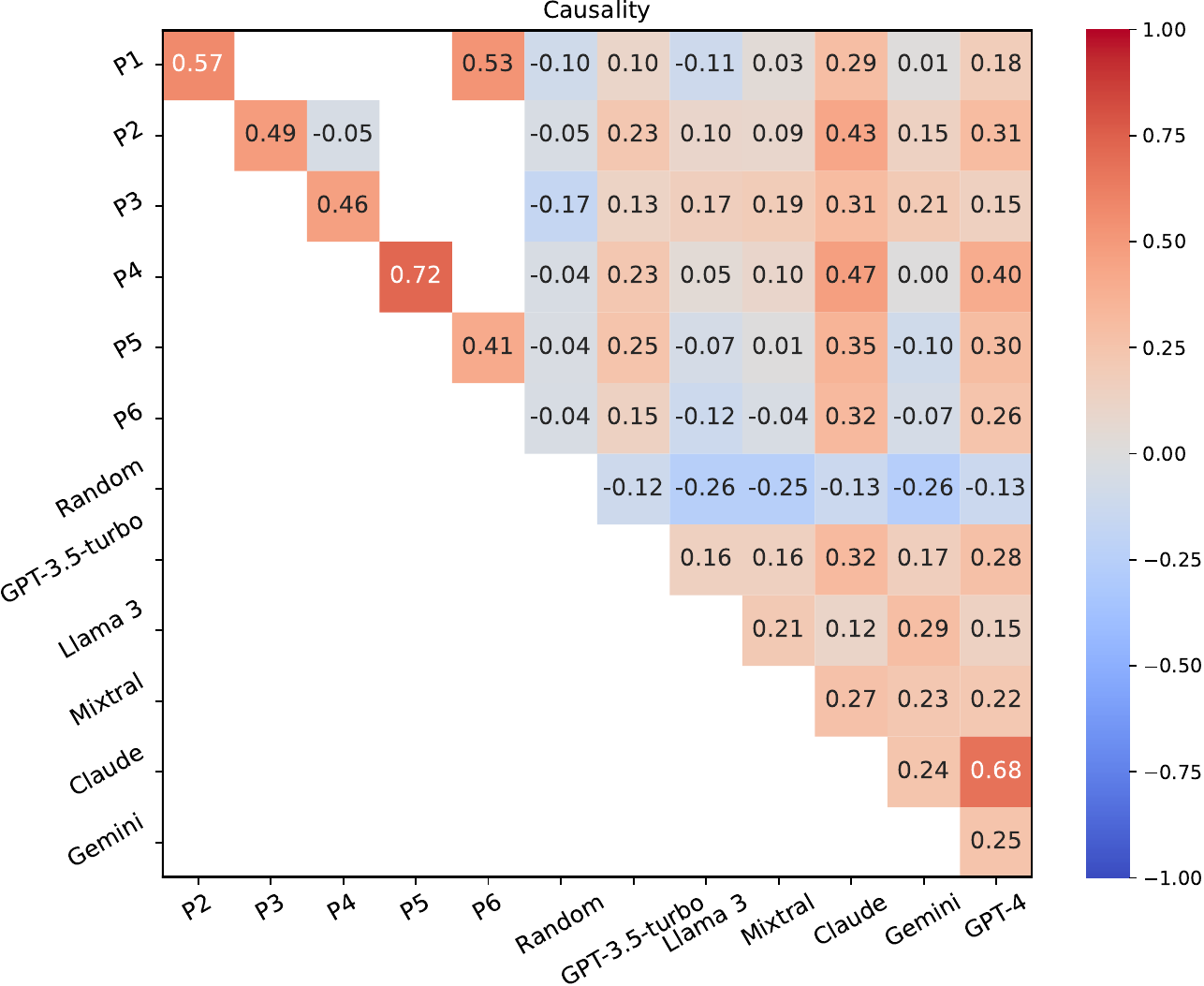}
    \caption{Inter-rater agreement for causality.}
   \label{fig:causal}
\end{figure}

\smallskip
\noindent{\underline{\em Causality}}
The models struggle more to detect causality, relative to the two prior tasks (Figure~\ref{fig:causal}). The mean human-model agreement (0.22) is much less than human-human (0.44). The mean model-model agreement (0.39) is less than 0.5. To summarize, we found lower level of agreement between human-model \emph{and} model-model for this task.

\begin{figure}[!t]
    \centering
    \includegraphics[width=0.85\columnwidth]{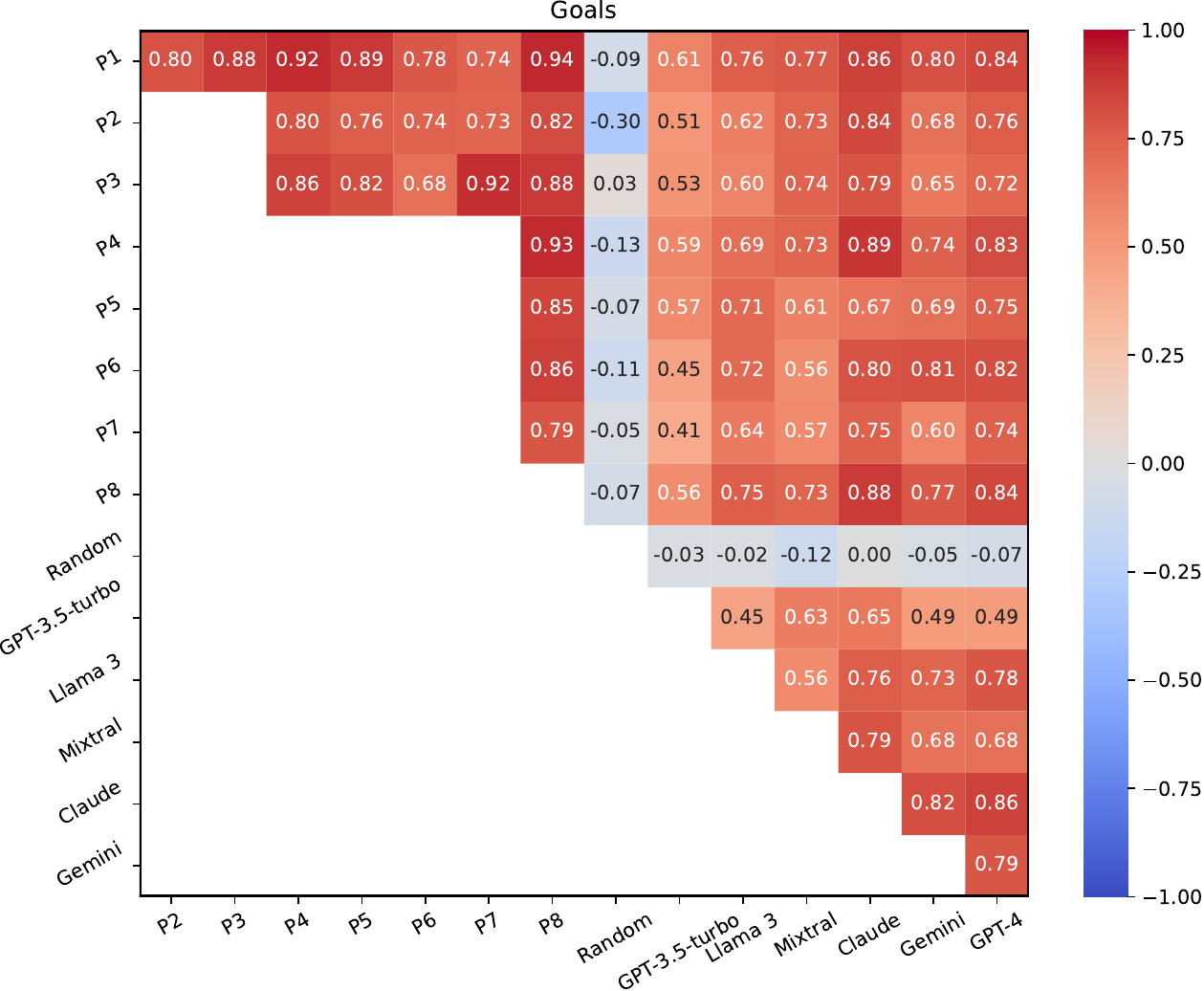}
    \caption{Inter-rater agreement for semantic similarity with evaluation criteria ``Goals''. Other criteria (``Operations'' and ``Effects'') give similar heatmaps (omitted due to space).}
   \label{fig:sesame}
\end{figure}

\smallskip
\noindent{\underline{\em Semantic Similarity}}
For the functional semantic similarity dataset in all categories (e.g., goals, operations, and effects), we have observed high agreement for all human-human (0.71-0.83), human-model (0.64-0.77), and model-model (0.69-0.83) pairs (see Table~\ref{tbl:sira-all} and Figure~\ref{fig:sesame}). We found the highest level of inter-rater agreement in all three pairings.

\begin{comment}
\begin{figure*}[ht]  % The 'figure*' environment ensures it spans across two columns
    \centering
    \begin{subfigure}[b]{0.30\textwidth}  % Each subfigure takes up 25% of the text width
        \centering
        \includegraphics[width=\linewidth]{Figures/Goals.pdf}  % Scale the image to fit the subfigure width
        \caption{Goals}
    \end{subfigure}%
    \begin{subfigure}[b]{0.30\textwidth}
        \centering
        \includegraphics[width=\linewidth]{Figures/Operations.pdf}
        \caption{Operations}
    \end{subfigure}%
    \begin{subfigure}[b]{0.30\textwidth}
        \centering
        \includegraphics[width=\linewidth]{Figures/Effects.pdf}
        \caption{Effects}
    \end{subfigure}%

    \caption{Inter-Rater Agreement for Different Semantic Similarity Evaluation Criteria.}
    \label{fig:sesame}
\end{figure*}
\end{comment}

\begin{figure}[!t]
    \centering
    \includegraphics[width=0.85\columnwidth]{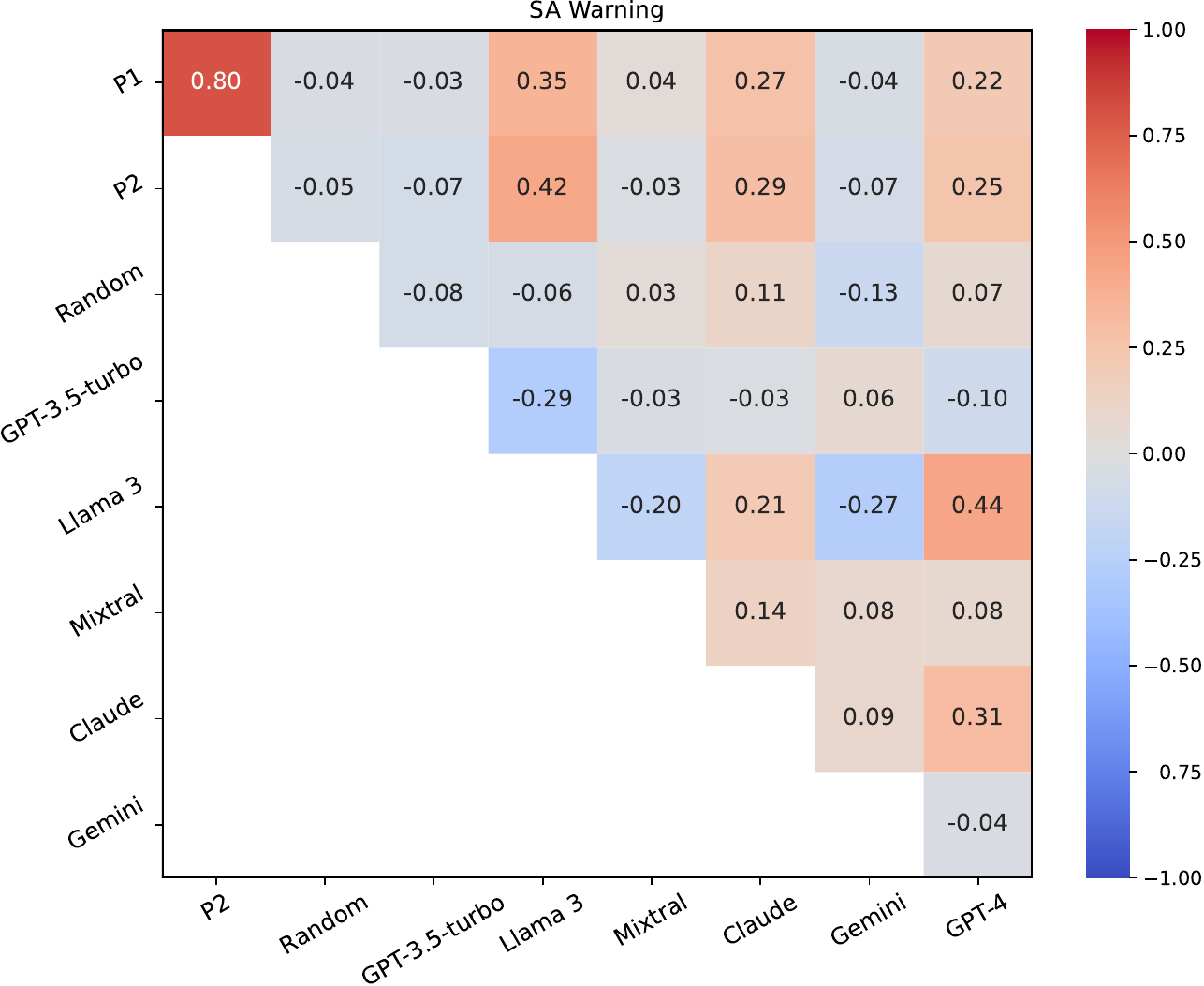}
    \caption{Inter-rater agreement for static analysis warnings.}
   \label{fig:sa}
\end{figure}

\smallskip
\noindent{\underline{\em Static Analysis Warning}}
For the static analysis warnings rating task, we have only two human raters, and they are in strong agreement (Krippendorff’s $\alpha$ 0.80). However, human-model (0.15) and model-model (0.12) agreements are low (Figure~\ref{fig:sa}). These findings suggest that LLMs cannot safely substitute
for human ratings in this task. 

We note that human-model agreement varies from task to task.
While it resembles human-human agreement values for some tasks (code summarization, name-value inconsistencies, semantic similarity), it is lower for other tasks (causal relation detection and static analysis warnings). For some tasks, all reviewers labeled all the samples (e.g., static analysis warning and name-value inconsistency). For code summarization, the annotation load was equally distributed among the annotators. In code summarization, each pair of humans had 105 overlapping annotations. However, for causality, the load was not equally distributed.  In some cases, there are no samples labeled by two individuals, and in those cases, the corresponding cell is empty.

\subsection{RQ2: LLMs' Applicability for a Specific Task}
\label{res_rq2}

\begin{figure}[!t]
    \centering
    \includegraphics[width=0.85\columnwidth]{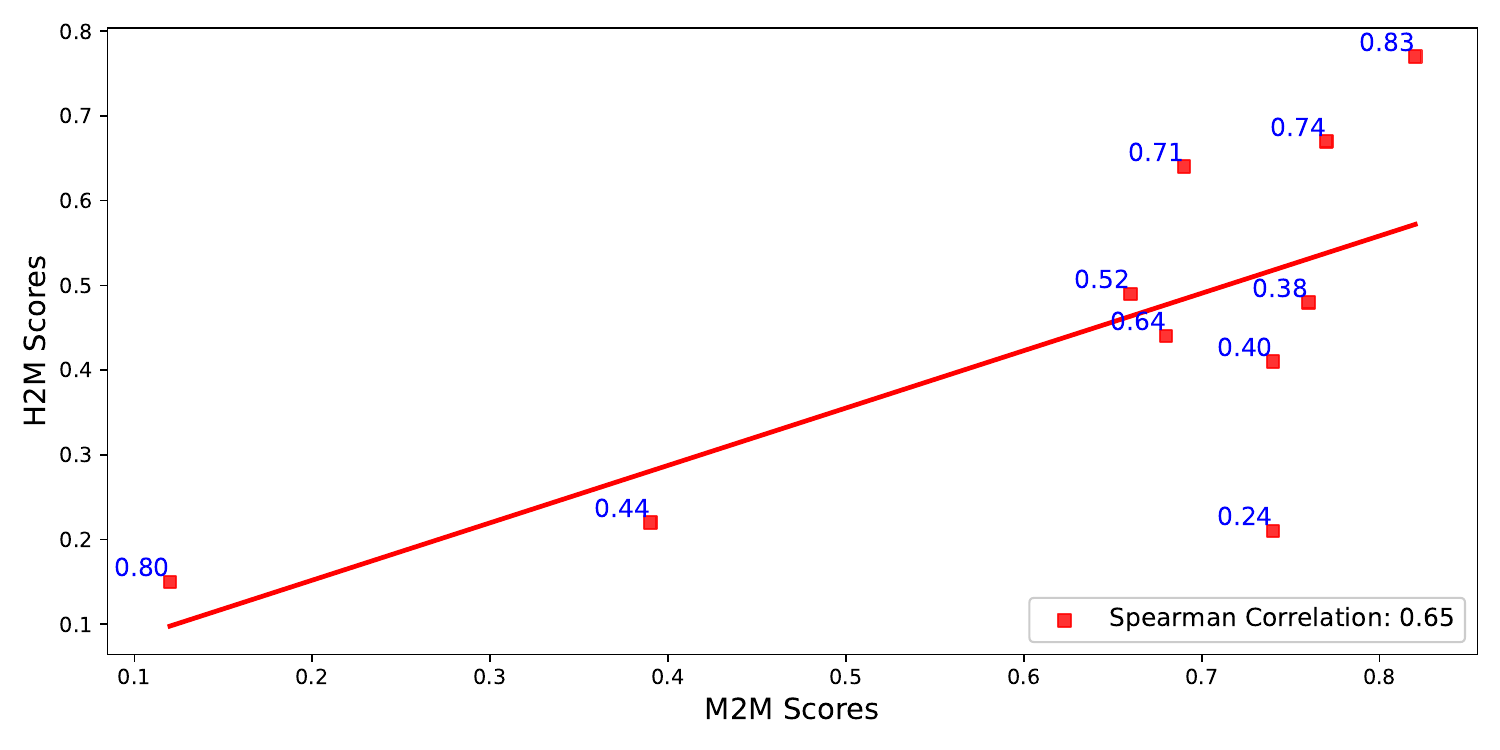}
    \caption{Human-model inter-rater agreements are positively correlated with model-model inter-rater agreements ($p<0.05$). Human-human agreements are shown in blue text.}
   \label{fig:scatter}
\end{figure}

We observed that for a majority of tasks, human-model inter-rater agreement is very similar to human-human agreement, while for some tasks, it is lower. What about model-model inter-rater agreement, which does not require any human effort? Figure~\ref{fig:scatter} shows that the mean model-model agreement of the top 3 models is positively correlated with the mean human-model agreement, with a Spearman correlation of 0.65 ($p < 0.05$). This suggests that higher model-model agreement is helpful in indicating good human-model agreement; this observation is potentially valuable in deciding whether to replace humans with models. There is an outlier (high model-model agreement but low human-model agreement) in the bottom-right corner, but for this case (code summarization conciseness), \emph{human-human agreement is also low} (0.24) suggesting that human ratings for this task \emph{per se}
are not consistent.
Our results suggest that if multiple LLMs reach similar solutions independently, then LLMs are likely suitable for the annotation task. 

\subsection{RQ3: LLMs' Applicability for a Specific Sample}
\label{res_rq3}
If we observe the tasks where the models and humans are reasonably agreeing with each other, the model-model agreements are quite high (0.66-0.82). For two datasets, causality detection and static analysis warning, the inter-model agreements are too low (0.39 and 0.12). Now we will see whether we can replace one human rating in each sample with a model.
We will discuss the results with respect to the GPT-4 model output from here on, because this model allows us to observe the output probability and is the best-performing model overall.

\begin{figure*}[ht]  % The 'figure*' environment ensures it spans across two columns
    \centering
    \begin{subfigure}[b]{0.50\textwidth}  % Each subfigure takes up 25% of the text width
        \centering
        \includegraphics[width=.85\linewidth]{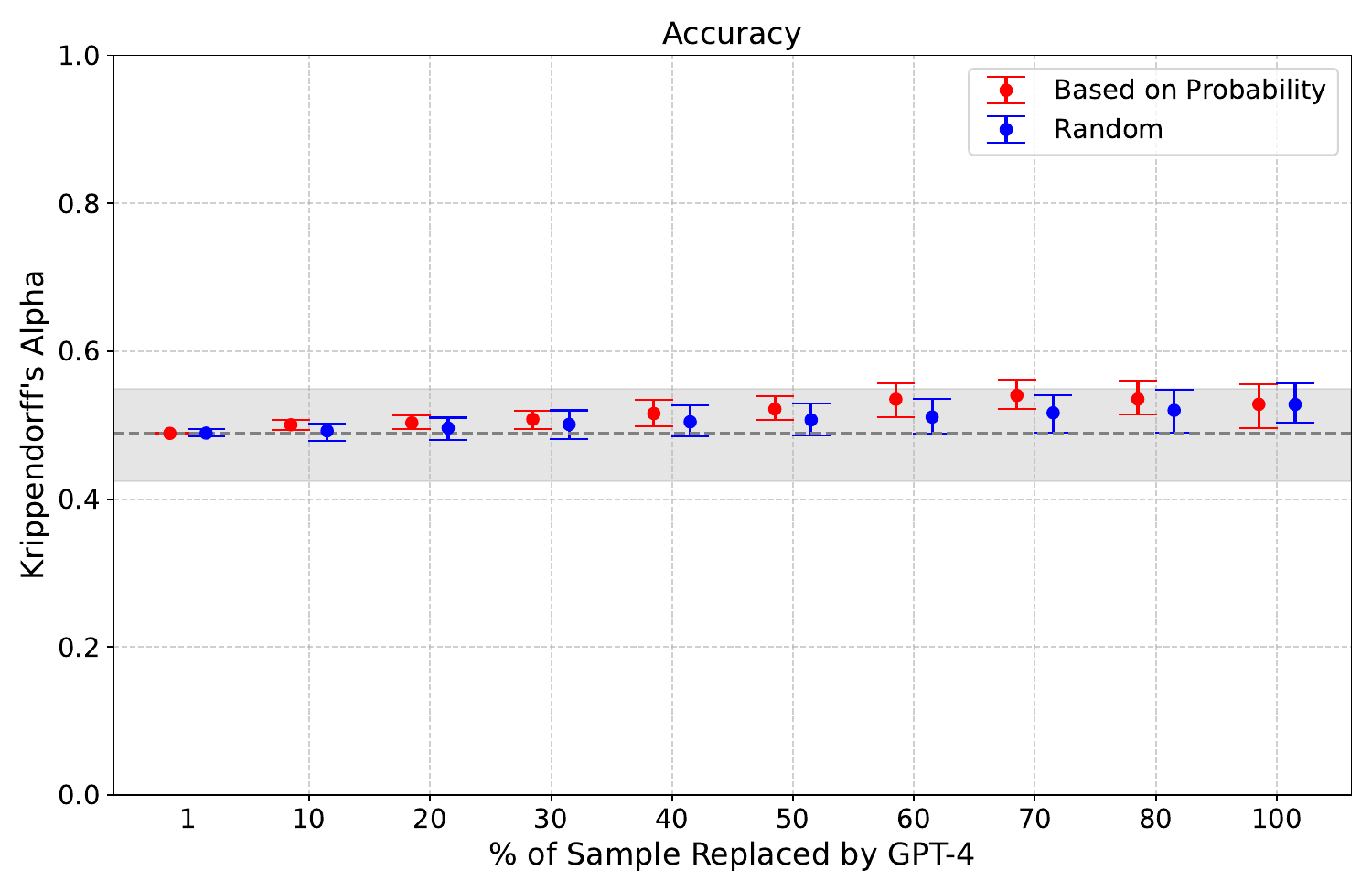}  % Scale the image to fit the subfigure width
        \caption{Accuracy}
    \end{subfigure}%
    \begin{subfigure}[b]{0.50\textwidth}
        \centering
        \includegraphics[width=.85\linewidth]{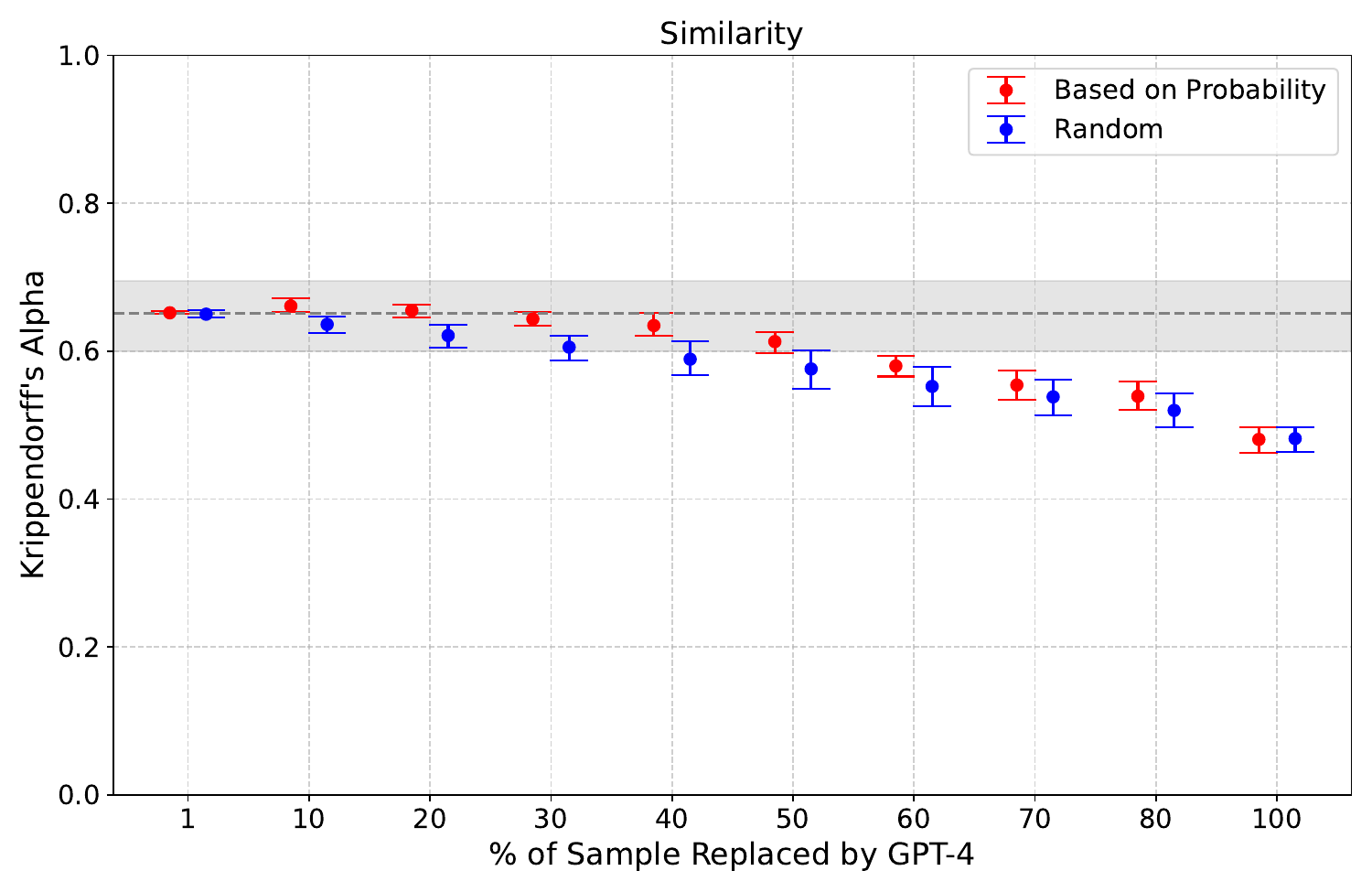}
        \caption{Similarity}
    \end{subfigure}
    \caption{Inter-rater agreement with increasing \% of samples from GPT-4 for code summarization accuracy and similarity. The dotted line indicates the inter-human agreement.}
    \label{cs-sample}
\end{figure*}

Figure~\ref{cs-sample}(a \& b) shows how inter-rater agreement changes with increasing fraction of samples from GPT-4 for code summarization accuracy and similarity. For adequacy and conciseness, we have similar plots to accuracy, so we have omitted them from the paper due to page constraints. Consider Figure~\ref{cs-sample}(a), where we gradually replace the human ratings with model ratings. On the x-axis, 10\% means we take 10\% of the samples and replace one randomly chosen rating with GPT-4 output, then observe the inter-rater agreement (now including 10\% model ratings).

We can choose the 10\% samples in two ways: based on GPT-4 output probability or randomly. The red lines show the results when we choose the samples based on probability, and the blue line indicate when samples are chosen randomly. After selecting the samples, we have three ratings for each sample. We randomly select a position, replace the rating with GPT-4 output, and calculate Krippendorff's $\alpha$ as in RQ1. We repeat the process 100 times for each selection criterion. 
For each data point on the x-axis, we present the mean value and 95\% confidence interval.
For human-human agreement, we present the inter-rater agreement with a dotted line and have drawn the confidence interval using bootstrapping by randomly selecting 50\% of the samples and repeating the process 1,000 times.

In general, the probability-based selection criterion works better than randomly selected samples. However, at 100\%, there is no difference between them because all the samples have been considered for replacement.
For the accuracy task of code summarization (Figure~\ref{cs-sample}), we find that the inter-rater agreement does not change at all, even when we replace one rating in all the samples. It slightly increases with the probability-based selection approach, the improvement is insignificant and within the 95\% confidence interval of human-human agreement.
For the similarity task (Figure~\ref{cs-sample}-b), we have seen that GPT-4 maintains inter-rater agreement up to 50\% of the samples while chosen based on probability, potentially saving 16.5\% of overall human effort.

\begin{figure}[!t]
    \centering
    \includegraphics[width=0.85\columnwidth]{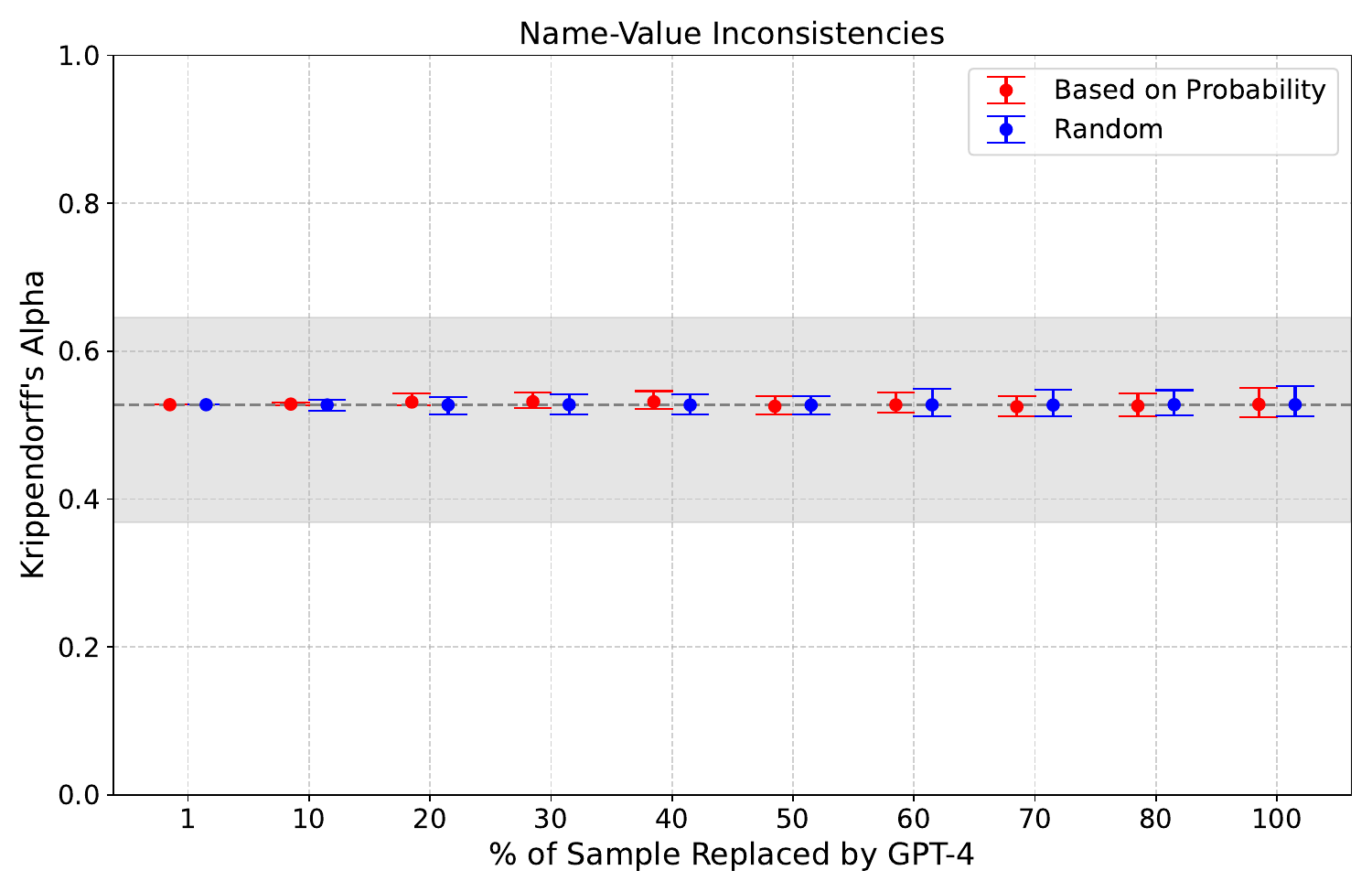}
    \caption{Inter-rater agreement with increasing \% of samples from GPT-4 for name-value inconsistencies.}
   \label{nalin-sample}
\end{figure}

\begin{comment}
\begin{figure}[!t]
    \centering
    \includegraphics[width=0.95\columnwidth]{Figures/goals_sample-crop.pdf}
    \caption{Inter-Rater Agreement with increasing \% of Samples from GPT-4 for Semantic Semilarity (Goals).}
   \label{sesame-sample}
\end{figure} 

\end{comment}

For name-value inconsistencies (Figure~\ref{nalin-sample}) and functional similarity (figure omitted for space constraints), we observed that we can replace one rating in all samples with GPT-4 and still maintain the same inter-rater agreement. Note that for name-value inconsistency, we have few samples. Therefore, the confidence interval of human-human agreement is much wider.

%Note that in Figure X, we only show the results for the criterion ``goals''. We omitted the results for ``operations'' and ``effects'' due to page constraints, but we observed similar plots for these criteria.  

\begin{figure}[!t]
    \centering
    \includegraphics[width=0.85\columnwidth]{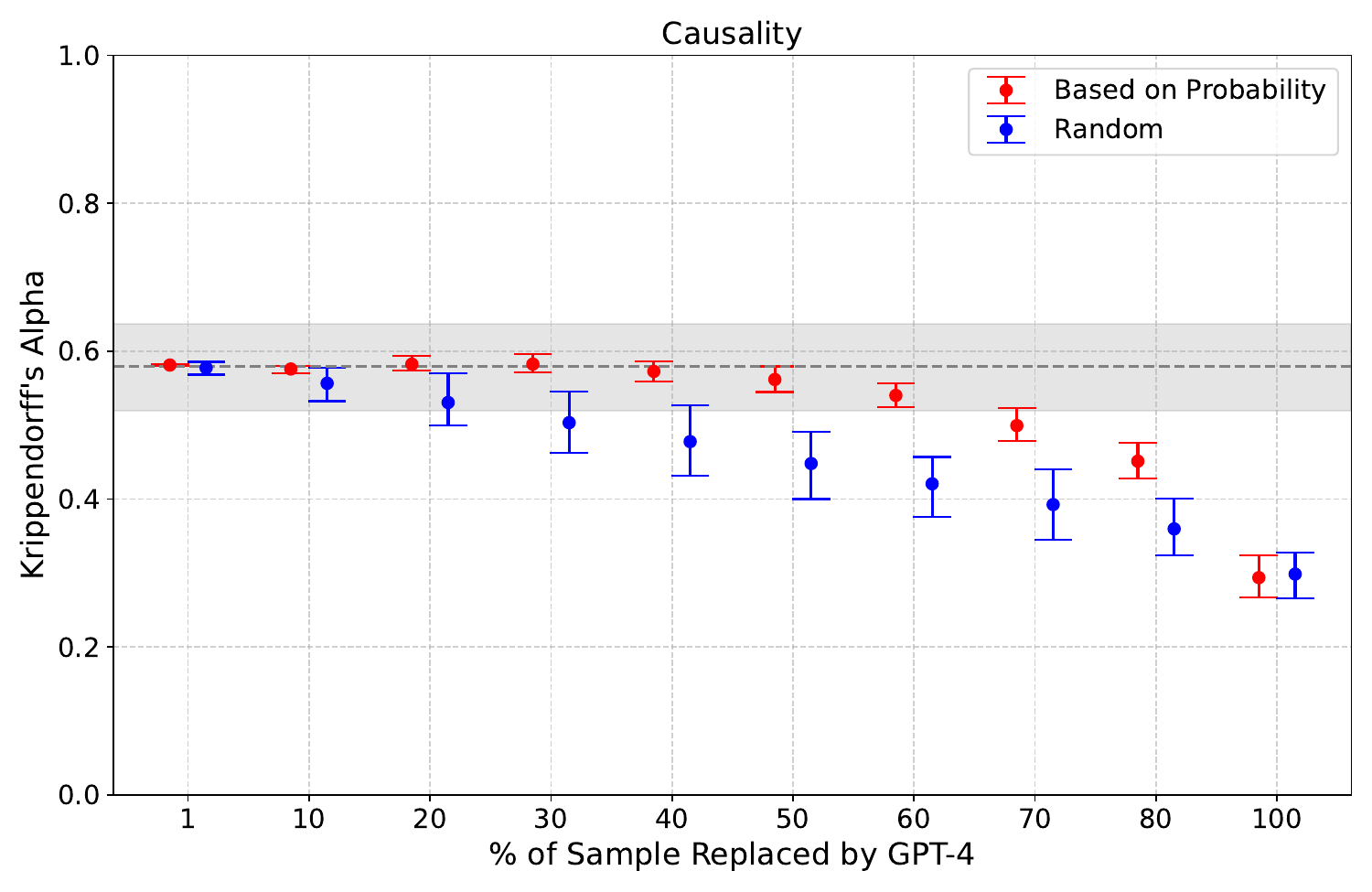}
    \caption{Inter-rater agreement with increasing \% of samples from GPT-4 for causality.}
   \label{causal-sample}
\end{figure}

\begin{figure}[!t]
    \centering
    \includegraphics[width=0.85\columnwidth]{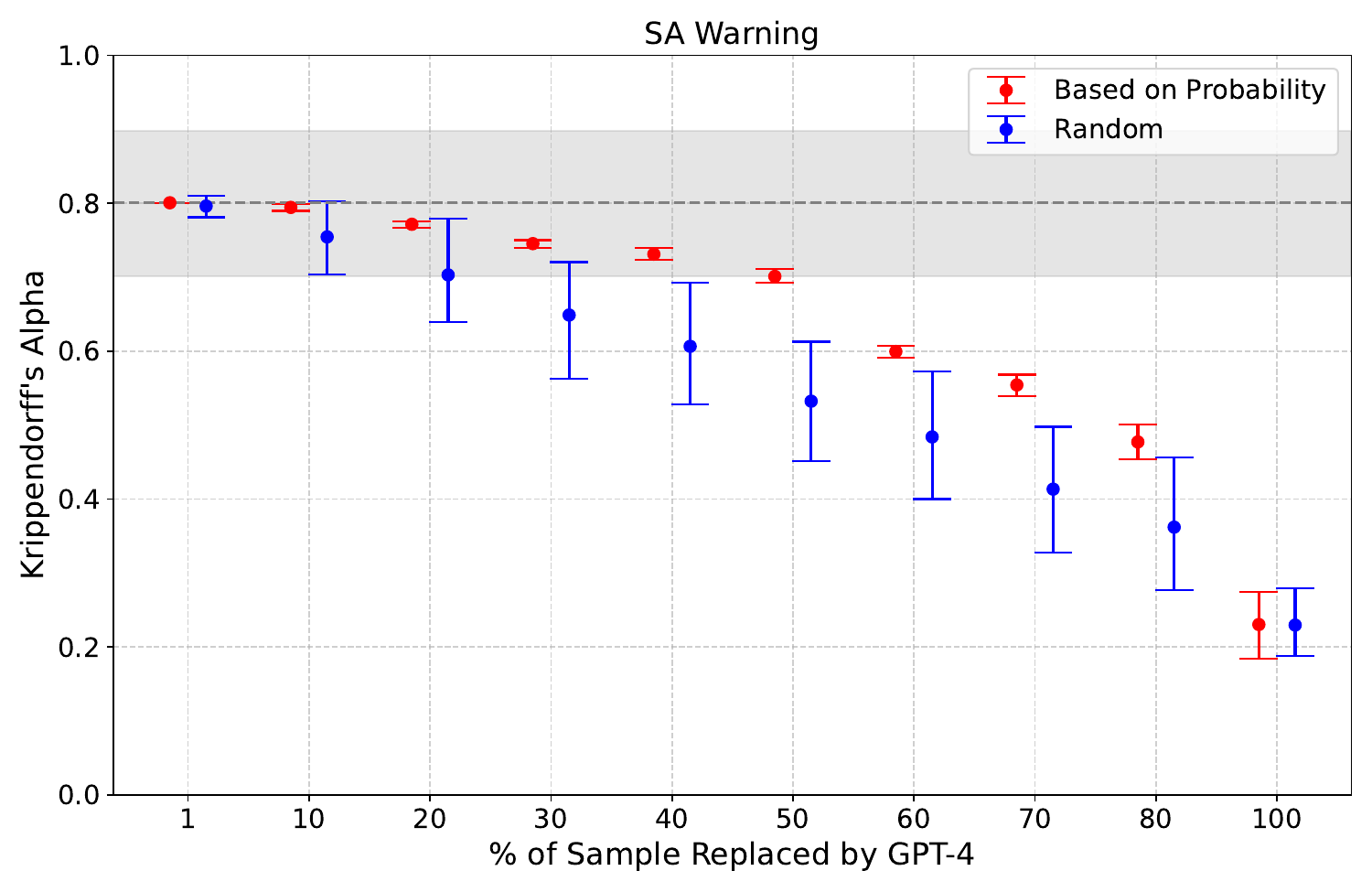}
    \caption{Inter-rater agreement with increasing \% of samples from GPT-4 for static analysis warnings.}
   \label{sa-sample}
\end{figure}

For causality and static analysis warning, we have low inter-model agreement (0.39 and 0.12). We have already discussed that we should expect low human-model agreement because they are correlated. Figures~\ref{causal-sample} \& \ref{sa-sample} show that the inter-rater agreement decreases as we increase the samples from the GPT-4 model, indicating that the model may not be applicable in this setup. However, using the confidence-based approach, we can still see that GPT-4 maintains similar inter-rater agreement, up to 60\% for the causality dataset and 50\% for the static analysis dataset, showing some promise for using GPT-4 for partial datasets. The confidence interval for these datasets up to that point completely overlaps with the confidence interval of the human-human rating.
High-probability samples are likely to be more correct or align better with human preferences compared to other samples.

\begin{table}[h]
%\caption{Effectiveness of multi-lingual fine-tuning for code summarization task}

\centering

\resizebox{.90\columnwidth}{!}{%
\begin{tabular}{lccc}

\hline
\multicolumn{1}{c}{Datasets}     & \begin{tabular}[c]{@{}c@{}}\#of Rating for \\ Each Sample\end{tabular} & \begin{tabular}[c]{@{}c@{}}\% Effort Saved \\ for One Rating\end{tabular} & \begin{tabular}[c]{@{}c@{}}\% Effort Saved \\ for Overall Process\end{tabular} \\ \hline
Accuracy (Code Summarization)    & 3                                                                      & 100\%                                                                    & 33\%                                                                          \\
Adequacy (Code Summarization)    & 3                                                                      & 100\%                                                                    & 33\%                                                                          \\
Conciseness (Code Summarization) & 3                                                                      & 100\%                                                                    & 33\%                                                                          \\
Similarity (Code Summarization)  & 3                                                                      & 50\%                                                                     & 16.5\%                                                                        \\
Nalin                            & 11                                                                     & 100\%                                                                    & 9\%                                                                           \\
Causality                        & 2                                                                      & 60\%                                                                     & 30\%                                                                          \\
Goals (Sesame)                   & 3                                                                      & 100\%                                                                    & 33\%                                                                          \\
Operations (Sesame)              & 3                                                                      & 100\%                                                                    & 33\%                                                                          \\
Effects (Sesame)                 & 3                                                                      & 100\%                                                                    & 33\%                                                                          \\
Static Analysis Warning          & 2                                                                      & 50\%                                                                     & 25\%   \\ \hline

\end{tabular}

}
\caption{\% of human effort saved for one rating and overall annotation process.}
\label{tbl:humaneffort}
\end{table}

\subsection{RQ4: Human Effort to Potentially Save}
\label{res_rq3}

By replacing one human rating with an LLM's answer for some fraction of all samples, as studied in RQ3, we can save some human effort.
The saved effort depends on two factors:
First, the number of ratings required per sample. As we replace only one human's rating for each sample, more required ratings per sample means less potential for saving human effort.
Second, the fraction of samples for which we replace a human rating with an LLM's rating.
To decide on this fraction, we use the results from RQ3 to determine the maximum fraction that keeps the resulting inter-rater agreement statistically indistinguishable from a human-only study.
For example, in Figure~\ref{cs-sample}-b, asking an LLM for help for up to 50\% of all samples keeps the agreement within the zone marked with gray background, but going beyond that fraction would change the inter-rater agreement in a statistically significant way.

Based on this reasoning, Table~\ref{tbl:humaneffort} summarizes the amount of effort that could potentially be saved.
For the accuracy task of code summarization, we can save 100\% of the effort to obtain one rating per sample, which is 33\% of the overall rating effort for this tasks.
For name-value inconsistencies, we can save only 9\% of the overall effort because there are 11 ratings per sample, whereas for the semantic similarity dataset, we can save up to 33\% of the effort because it involves only three ratings per sample.
Note that we ignore the few-shot labeling effort (which is 3-4 samples only) for this estimation in Table~\ref{tbl:humaneffort}.
Overall, for seven of the ten tasks,  we can safely replace one human rater with a model.

\section{Discussion}
\label{sec:discussion}

\begin{figure*}[!t]
    \centering
\includegraphics[width=.80\textwidth, trim=0cm 7cm 0cm .5cm, clip]{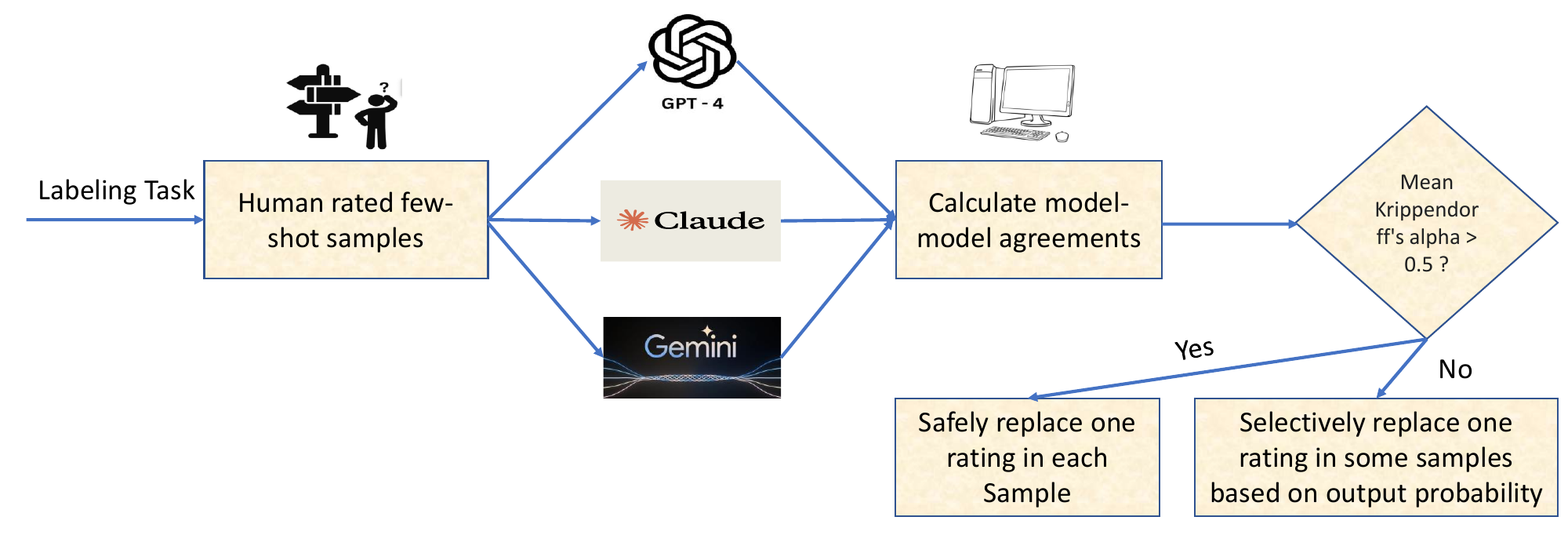}
    \caption{Steps to decide human rating replaceability.}
   \label{fig:steps}
\end{figure*} 

\smallskip
\noindent{\underline{\em Deciding Whether and When to Ask an LLM}} 
Our findings suggest that we can replace one human rating by a model for a significant number of samples (50\%-100\%). However, it appears risky to fully replace all humans, even though the probability of the model output 
at least somewhat indicates the quality of annotations.
Based on our findings, we propose the process as given in Figure~\ref{fig:steps}, which involves two steps: 
1) Create few-shot examples (3--4 in our setup) to query multiple strong LLMs with all samples, and compute the model-model agreement.
2) If the agreement is high ($>0.5$ based on our data), one can safely replace one human rating per sample with an LLM-provided answer. Otherwise, selectively replace one rating only for those samples where the LLM gives a high-confidence answer.
Beyond saving human effort, our findings can also help extend a dataset by automatically labeling additional samples.
We should emphasize that replacing more than one human can inflate inter-rater agreement because model-model agreements are much higher. This may not fulfill or reflect the original goals of the data annotations.

% For code summarization similarity, we can safely replace up to 50\% of the samples; the inter-rater agreement goes down after that. However, the decline is less severe relative to datasets where we got $< 0.50$ mean model-model agreement (causality and static analysis warning). Therefore, a mean model-model agreement $>0.50$ suggests that one human rating in each sample can be automated away with GPT-4.

\begin{figure}[!t]
    \centering
    \includegraphics[width=0.95\columnwidth]{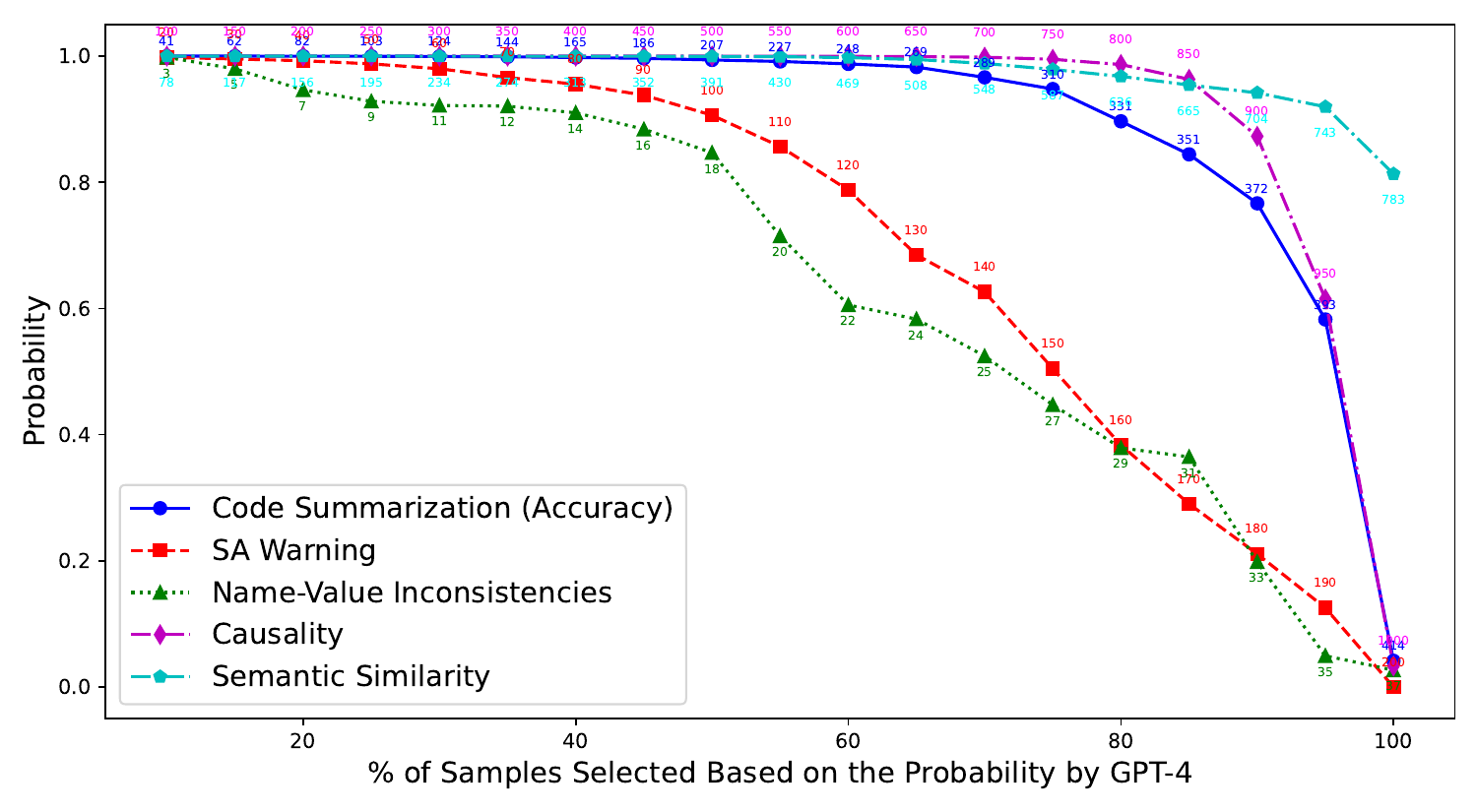}
    \caption{Change of output probability (by GPT-4) with \% of samples sorted based on GPT-4 model's output probability. Some datasets were omitted for clarity of the plot.}
   \label{fig:sample-prob}
\end{figure}

\smallskip
\noindent{\underline{\em Confidence Distribution over Samples?}}
We did find that picking the samples based on model confidence yields better results, even when model-model agreement is low, for almost 50\% of the data. However, the question remains: what should be our cutoff confidence? Figure~\ref{fig:sample-prob}, 
which orders fraction of the sample, at decreasing confidence levels, shows 
that GPT-4 is usually quite confident with its solutions. However, after a certain point, the probability starts to decrease. For causality and static analysis warning, we can see that for the first 50\%-60\% of samples, the model maintains a high output probability, around 0.80. Thus, the probability cutoff point can be computed when we have low model-model agreement.

\noindent{\underline{\em Deciding Whether All Humans Are Replaceable}} 
In RQ3, we asked if one rating in all samples can be replaced by an LLM.  Can we replace \emph{all} humans, but just for \emph{specific} samples? First, we clarify when we can replace all the humans for a sample. Note that there are very few samples where all annotators would agree. For name-value inconsistencies, no sample exists where all annotators agree on a particular rating. Therefore, we will use majority voting to decide whether all humans are replaceable for a sample or not. For a dataset with 2 annotators, the model needs to agree with both annotators, and for a dataset with 11 annotators, the model needs to agree with at least 6 annotators to replace all humans. 
%Based on this criterion, we have a sample ROC-AUC curve~\ref{fig:csauc} for code summarization accuracy for 4 models, where we could obtain the output probability from the model. Table~\ref{tbl:auc} shows the number of samples, number of true positives, and AUC for all datasets. Probabilities from other models (e.g., GPT-3.5-turbo, Llama3, Mixtral) sometimes show more predictive strength because of high AUC but have a low number of true positive samples. Therefore, we opt for GPT-4 to guide the discussion.
%AUCs for GPT-4 model range from 0.59-0.89. From these values, we found that the model's probabilities have some predictive strength but are far from perfect for most of the datasets. At the thresholds, positive and negative samples are not clearly distinguishable. We can replace humans for selective samples, but there is always some risk associated with it.

\begin{figure}[!t]
    \centering
    \includegraphics[width=.90\columnwidth]{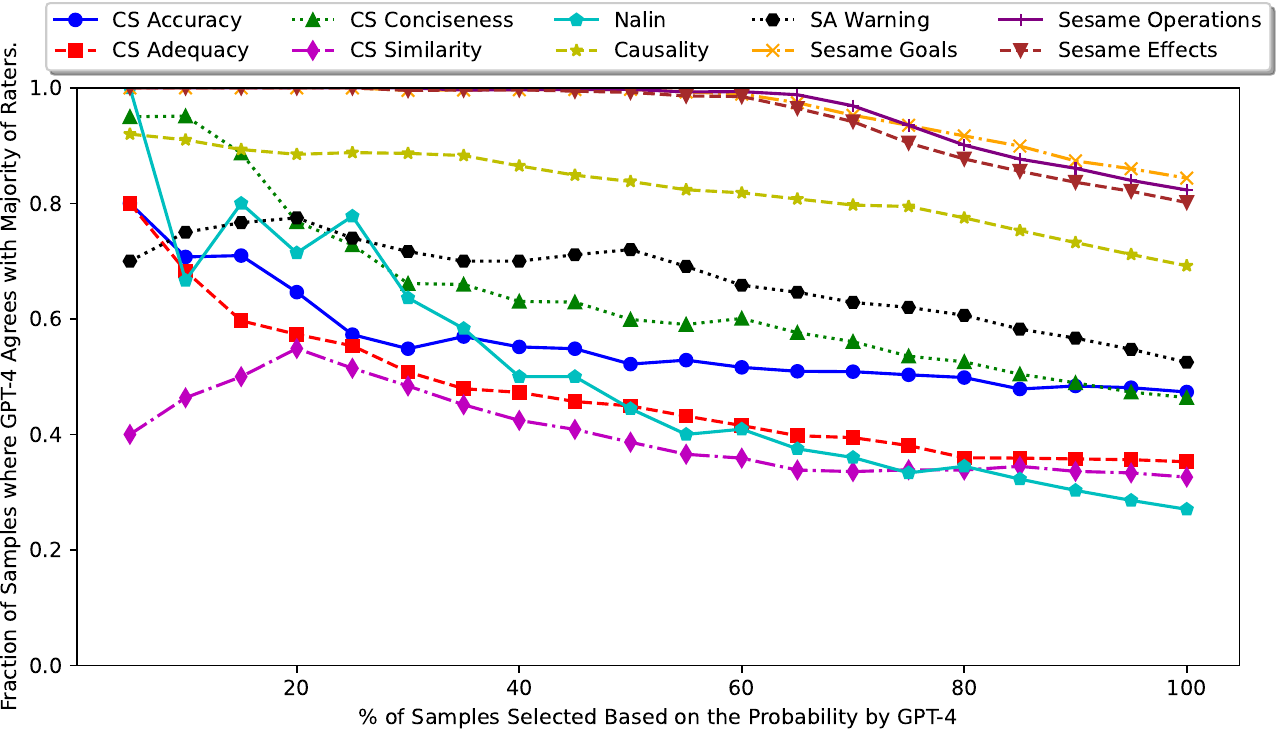}
    \caption{Change of fraction of samples where GPT-4 agrees with majority voting for \% of samples sorted based on GPT-4 model's output probability.}
   \label{fig:sample-major}
\end{figure}

Figure~\ref{fig:sample-major} shows that samples with higher GPT-4 output probability are more likely to agree with the majority of annotations; however, even with 10\% of the samples, there is some possibility of error. Apart from semantic similarity, across all datasets, we can’t find any probability split where the model consistently agrees with majority voting. Therefore, we conclude that we cannot replace all the humans for a sample with the current models. We can replace humans for selective samples, but there is some risk associated with it. More study is required for this; we leave it for future research.

%\smallskip
%\noindent{\underline{\em Limitations}} 
%We consider only 10 tasks datasets and 6 models. 
%It is quite difficult to find publicly available datasets where the ratings from individual annotators are still available; many artifacts contain only the final annotation after resolving disagreements.
%Even though we see pretty consistent results across the datasets and models, finding may vary with others. 
%Only Krippendorff’s $\alpha$ was applicable to all studied datasets. 
%Cohen's $\kappa$ is another popular measure but is applicable to only two raters.
%Using few-shot learning means that some human effort is required initially.
%However, labeling 3--4 samples is much cheaper compared to labeling all the samples. 
% We used the same prompts used by the original study and the labels we found in the repository, except for changing the negative
% sense in some instructions to positive (See \autoref{ref:neg}).  

\smallskip
\noindent{\underline{\em Further Investigations}} 
It is now well-established that few-shot learning performs better than zero-shot learning. We also tried zero-shot learning in our preliminary stage and found that the human-model agreement was not satisfactory, even for datasets where few-shot learning performed very well. We tried repeated sampling (at higher temperature): the models mostly generate the same samples. We could improve the multi-sample technique by using chain-of-thought~\cite{wei2022chain} or self-consistency~\cite{wang2022self}, but we do not have the original thoughts used by the original human annotators. In future studies, we strongly recommend collecting human thoughts, which could be used for few-shotting.

%Why few-shot instead of zero shot?

%Why not single model with multiple sample?

%Why not self ask?

%\noindent{\underline{\em Based on 5 datasets}} 

%\noindent{\underline{\em One inter-rater agreement}}

%\noindent{\underline{\em Depends on model}}

%\noindent\underline{\bf Threats}:
%\prem{we don't look at demographics/bias}

\section{Threats to Validity}

We consider only 10 human-annotation tasks and 6 models. 
It is difficult to find publicly available datasets where the ratings from individual annotators are still available; many artifacts contain only the final annotation after resolving disagreements.
Even though we see fairly consistent results across the datasets and models, finding may vary with others. 
Only categorical Krippendorff’s $\alpha$ was applicable to all studied datasets. For some of our tasks, ordinal Krippendorff’s $\alpha$ is also applicable (e.g., code summarization). However, we observe negligible change from using ordinal Krippendorff’s $\alpha$.  

Cohen's $\kappa$ is another popular measure \emph{but is applicable to only two raters}. We  calculated Cohen's $\kappa$ for each human and model pairs. Although we observe lower values with Cohen's $\kappa$, the difference between human-human and human-model agreements remains similar. We share the detailed results in the supplementary material.
We proposed a 0.50 inter-model threshold based on our experiments; this may not always work (e.g., in scenarios where typical agreement values are higher or lower).
%Cohen's $\kappa$). 
Another possibility (besides using an inter-model agreement threshold), is to manually label a subset of the data; if human-human and human-model IRAs, are similar in this subset, the model could replace one human.

Few-shot learning requires some human effort initially, to get the
``shots". 
However, labeling 3--4 samples is much cheaper compared to labeling all the samples. Also, 
in few-shot learning, we present the model with one randomly chosen example from each category to prevent bias. However, outcomes may depend on the set of few-shot samples used. 

Although all of our datasets are relatively recent, some of them might be part of the training data of some of the LLMs we evaluate.
As our methodology performs several pre-processing steps, such as function body accumulation (for the semantic similarity task) and diff generation (for the static analysis warnings), it is highly unlikely that the model has memorized the exact results.
However, we acknowledge that some bias may exist due to the model’s exposure to the underlying GitHub repositories.

\section{Related Work}

\noindent{\underline{\em ML and LLMs in Software Engineering}}
Learning-based approaches and LLMs have been applied to various software development tasks~\cite{NeuralSoftwareAnalysis}, including
code completion~\cite{Chen2021,Nie2023,arXiv2024_De-Hallucinator},
test generation~\cite{Lemieux2023,schafer2023empirical,Ryan2024,Pizzorno2024},
fuzzing~\cite{Deng2023,icse2024-Fuzz4All},
automated repair~\cite{Chen2021d,Xia2023a,Ye2024,Silva2024,Hossain2024} and coding agents~\cite{Bouzenia2024,Yang2024a,Zhang2024a,Tao2024},
agents for other software engineering tasks~\cite{Bouzenia2024a},
and type prediction~\cite{Hellendoorn2018,icse2019,fse2020,Allamanis2020}.
We follow this trend, but deviate by not targeting a specific task, but by studying the potential for automating human-subject (manual) evaluations, using LLMs.

\smallskip
\noindent{\underline{\em LLMs to Complement or Replace Human Annotators}}
Recent and concurrent work explores the idea of complementing or replacing human annotators with LLMs in domains beyond software engineering.
Work in natural language processing (NLP) compares LLMs and human participants, typically recruited via crowd-working platforms, and report mixed results.
While some studies show that adding LLM labels can improve the aggregated labels of a dataset~\cite{DBLP:journals/corr/abs-2401-09760} and sometimes outperform crowd-workers~\cite{DBLP:conf/chi/HeHDRH24}, others note that LLMs fail to accurately represent differences between demographic groups~\cite{DBLP:journals/corr/abs-2402-01908}.
Others suggest to use LLMs to perform sub-tasks in a crowd-sourcing pipeline~\cite{DBLP:journals/corr/abs-2307-10168}.
Bavaresco et al.~\cite{bavaresco2024llms} propose a benchmark of NLP datasets to compare human and LLM performance.
We envision our work to serve as such a benchmark in software engineering.
To the best of our knowledge, we are the first to carefully evaluate automated LLM-annotations for software engineering artifacts, and 
%Moreover, we 
propose actionable guidelines for when and how to use LLMs to create annotations.

\smallskip
\noindent{\underline{\em Collaborative Human-LLM Annotation}}
Our suggested workflow (Section~\ref{sec:discussion}) relates to work on human-LLM collaboration for data labeling~\cite{DBLP:conf/eacl/KimMCRZ24}.
Some papers~\cite{DBLP:conf/chi/Wang0RMM24,goel2023llms} suggest first querying an LLM, and asking humans to refine some of the LLM-provided labels.
Instead, we find that inter-model agreement and LLM output probabilities provide an effective way of deciding which labeling tasks can be safely delegated to a model.

% \cite{DBLP:conf/chi/Wang0RMM24}: methodology for combining LLMs and humans on NLP labeling tasks; follows an iterative process, where humans verify LLM labels and (if quality is low) re-annotate data; instead, our process is about deciding on a per-task or per-instance level whether to use LLMs or humans

% \cite{goel2023llms}: LLMs to annotate medical texts; proposed methodology for combining human and LLM annotations, where humans refine LLM annotations

\begin{comment}
\smallskip
\noindent{\underline{\em LLMs to Automate Research Tasks}}
Our work fits into a broader stream of work on partially automating particular sub-tasks of performing academic research.
Liang et al.~\cite{Liang2024a} study to what extent LLMs can replicate empirical software engineering research by determining assumptions made in a study and by writing data processing code.
While we share the idea of using LLMs to automate a research task, we focus on a different kind of task.
Other work proposes to let LLMs generate open-ended questionnaire responses for HCI-style evaluations~\cite{DBLP:conf/chi/HamalainenTK23}, 
and shows a set of LLMs can reproduce human-study results in economics, psycholinguistics, and social psychology~\cite{aher2023using}.
We expect to see more work on delegating sub-tasks of conducting research to LLMs, e.g., for performing human-intensive evaluations and editing scientific texts.
\end{comment}

\smallskip
\noindent{\underline{\em LLMs for Data Synthesis and Assessment}}
Beyond using LLMs to partially automate evaluations that would otherwise be performed only by humans, others propose to use LLMs to synthesize additional data to train or fine-tune models for text annotation tasks~\cite{pangakis2023automated,DBLP:conf/emnlp/LiZL023}.
There is a recent related survey~\cite{DBLP:journals/corr/abs-2402-13446}.
A key difference from our work is that training and fine-tuning datasets must be large-scale, but some amount of noise is acceptable, whereas annotations in human-subject studies are typically of smaller scale, but should have high confidence.
LLMs have also been proposed as judges of outputs generated by other LLMs, either with a single LLM as the judge~\cite{Zheng2024} or with a set of LLMs that discuss until reaching an agreement~\cite{DBLP:journals/corr/abs-2308-07201}.
These efforts are primarily about determining human preference, i.e., tasks where different humans may legitimately disagree, whereas many annotation tasks in software engineering have an objectively correct answer that humans can eventually agree upon.

\section{Conclusion}
In this paper, we investigate if  LLM responses can substitute for human raters in software engineering annotation tasks,
and how such substitutions may affect inter-rater agreement. We find that human-model agreements can be fairly consistent
with human-agreements, for the majority of the settings we studied. We also find model-model agreement is a good indication
of human-model agreement, suggesting that when powerful models trained on giant human corpora agree with each other, 
they tend to agree also with humans. Finally, we find that an LLM's confidence (output probability) for a given sample
output is a good indication of whether LLM output agrees with majority human rating, for that sample. 
We caution that the paper considers only discrete (multiple choice) LLM responses, not free-form annotations; furthermore, we have not
studied issues of model output bias, nor issues of demographics.

Our scripts and datasets are publicly available: 
\url{https://zenodo.org/doi/10.5281/zenodo.13146386}

\section*{Acknowledgments}
This work was partially supported by the National Science Foundation under CISE SHF MEDIUM 2107592, the European Research Council (ERC, grant agreements 851895 and 101155832) and by the German Research Foundation within the ConcSys, DeMoCo, and QPTest projects.

\bibliographystyle{IEEEtran}
\bibliography{ref.bib,referencesMP}

%\newpage

%\input{supplementary}
\end{document}